\documentclass[aps,pra,floatfix,citeautoscript,superscriptaddress,reprint]{revtex4-1}
\usepackage{amsmath}
\usepackage{amssymb}
\usepackage{graphicx}
\usepackage{bm}
\usepackage{color}
\usepackage{times}
\usepackage{verbatim}
\usepackage[none]{hyphenat}

\newcommand{\bea}{\begin{eqnarray}}
\newcommand{\eea}{\end{eqnarray}}
\newcommand{\be}{\begin{equation}}
\newcommand{\ee}{\end{equation}}
\newcommand{\Ha}{\hat{\mathcal{H}}}

\newcommand{\vf}{v_F}
\newcommand{\pmas}{p_{+}}
\newcommand{\pmenos}{p_{-}}

\begin{document}
%%%%%%%%%%%%%%%%%%%%%%%%%%%%%%%%%%%%%%%%%%%%%%%%%%%%%%%%%%
\title{Anomalous Goos-H\"anchen shift in the Floquet scattering of Dirac fermions}
%\title{Floquet scattering of Dirac fermions from an irradiated region:  anomalous Goos-H\"anchen shift}
\author{A. Huam{\'{a}}n}
\affiliation{Centro At{\'{o}}mico Bariloche and Instituto Balseiro,
Comisi\'on Nacional de Energ\'{\i}a At\'omica (CNEA)--Universidad Nacional de Cuyo (UNCUYO), 8400 Bariloche, Argentina}
\affiliation{Instituto de Nanociencia y Nanotecnolog\'{i}a (INN), Consejo Nacional de Investigaciones Cient\'{\i}ficas y T\'ecnicas (CONICET)--CNEA, 8400 Bariloche, Argentina}
\author{Gonzalo Usaj}
\affiliation{Centro At{\'{o}}mico Bariloche and Instituto Balseiro,
Comisi\'on Nacional de Energ\'{\i}a At\'omica (CNEA)--Universidad Nacional de Cuyo (UNCUYO), 8400 Bariloche, Argentina}
\affiliation{Instituto de Nanociencia y Nanotecnolog\'{i}a (INN), Consejo Nacional de Investigaciones Cient\'{\i}ficas y T\'ecnicas (CONICET)--CNEA, 8400 Bariloche, Argentina}
%%%%%%%%%%%%%%%%%%%%%%%%%%%%%%%%%%%%%%%%%%%%%%%%%%%%%%%%%%
%%%%%%%%%%%%%%%%%%%%%%%%%%%%%%%%%%%%%%%%%%%%%%%%%%%%%%%%%%
\begin{abstract}
We study the inelastic scattering of two-dimensional massless Dirac fermions by an inhomogeneous time-dependent driving field. As a physical realization we consider a monolayer graphene normally illuminated with a circularly polarized laser of frequency $\Omega$ in a given region. The interaction Hamiltonian introduced by the laser, being periodic in time, can be treated with the Floquet method which naturally leads to a multi-channel scattering problem. We analyze planar and circular geometries of the interface separating the irradiated and non-irradiated regions and find that there is an anomalous Goos-Hänchen shift in the inelastic channel. The latter is independent of the amplitude of the driving while its sign is determined by the polarization of the laser field. We related this shift with the appearance of topological edge states between two illuminated regions of opposite chiralities.
\end{abstract}
%%%%%%%%%%%%%%%%%%%%%%%%%%%%%%%%%%%%%%%%%%%%%%%%%%%%%%%%%%

\date{\today}
\maketitle

%%%%%%%%%%%%%%%%%%%%%%%%%%%%%%%%%%%%%%%%%%%%%%%%%%%%%%%%%%
\section{Introduction}
%%%%%%%%%%%%%%%%%%%%%%%%%%%%%%%%%%%%%%%%%%%%%%%%%%%%%%%%%%
Physical system subject to the action of time dependent periodic potentials present a variability of interesting phenomena, which have recently captured the attention of the research community as they might provide new routes for electronic and optoelectronic devices or lead to the observation of novel phases of matter, as for example Floquet topological insulators (FTI)~\cite{Oka2009,Kitagawa2010,Lindner2011,Rudner2013} or time crystals~\cite{Wilczek2012,Watanabe2015,Khemani2016,Else2016,Zhang2017}. The former, in particular, are a prominent example where driving  an otherwise ordinary material leads to the generation of light-induced topological properties~\cite{Kitagawa2011,Rudner2013}. Very much as  ordinary topological insulators~\cite{Kane2005,Koenig2007,Hsieh2008,Hasan2010,Ando2013}, FTI have a bulk gap in their non-equilibrium band structure (quasienergy spectrum) while the resulting Floquet-Bloch bands are characterized by non-trivial topological invariants~\cite{Kitagawa2010,Rudner2013,Carpentier2015,Perez-Piskunow2015,Nathan2015}. In addition, they host chiral/helical states at the sample boundaries \cite{Kitagawa2011,Perez-Piskunow2014,Usaj2014a}. Their properties have been extensively discussed in many different contexts, ranging from condensed matter systems to artificial optical and sound lattices or even cold atom systems~\cite{Gu2011,Calvo2011,Dora2012,Wang2013a,Rechtsman2013,Ezawa2013,FoaTorres2014,Goldman2014,Choudhury2014,
DAlessio2014,Dehghani2014,Liu2014,Kundu2014,DalLago2015,Seetharam2015,Iadecola2015,Dehghani2015,Sentef2015,Titum2015a,Farrell2015,Lovey2016,PeraltaGavensky2016,Lindner2017,Kundu2017,PeraltaGavensky2018a}.
Yet, the problem of inelastic  scattering of an impinging  particle upon an irradiated region with topological features has received much less attention.

Many analogies exist between the scattering of an electron beam from  electrostatic potentials in  $2$D electron gases and the one of a light beam from  an interface between two media of  different refraction index. In condensed matter physics they have been recognized and exploited long ago as for instance in the early beginning of quantum transport in mesoscopic heteroestructures~\cite{Beenakker1989,vanHouten1991,Spector1992}. Very recently, the analogy has been pushed even further, to the realm of metamaterials~\cite{Marques2007}, with the  proposal to construct Veselago lenses using $pn$ junctions in graphene \cite{Cheianov2007,Chen2013} to effectively built a negative refraction index for electrons. 
More recently, Beenakker \textit{et al}~\cite{Beenakker2009} have shown that quantum transport on a $pnp$ channel in graphene is affected by the presence of a subtle effect related to the scattering of an electron beam at the junction interface: the Goos-H\"anchen (GH) shift~\cite{Goos1947}. This is a  very well known effect in optics~\cite{Hentschel2002,Schomerus2006} that appears in the case of total reflection and corresponds to a lateral shift of the reflected beam of a magnitude comparable to the wavelength $\lambda_0$ due to interference effects. 
%While in a naive  semiclassical picture it can be accounted for as the displacing of the beam as it evanescently penetrates  inside the forbidden region, it is a purely  interference effect  that turns out to be proportional to the inverse $\lambda_0$.

In this work, we analyze the inelastic scattering of Dirac fermions from an irradiated region by considering both a planar and a circular interface. As the circularly polarized laser field of frequency $\Omega$ opens a {\it dynamical gap} in the Floquet spectrum of that region at a quasienergy $\hbar\Omega/2$~\cite{Calvo2011}, we take the energy $\varepsilon$ of the incident particles to be inside  such gap ($\varepsilon\sim\hbar\Omega/2$) so that only evanescent states penetrate the irradiated region. Then, we study the cases of an incident plane wave and a narrow beam  and discuss the GH shift.
Due to the topological character of irradiated region one can anticipate that the GH shift might present some features that are qualitatively different from those observed without the time dependent potential~\cite{Cserti2007}. We show that this is the case, and that the GH shift is not only different from zero at normal incidence but also its sign depends only on the direction of the laser's polarization. We interpret this as a consequence of the presence of a chiral current at the interface.

The rest of the paper is organized as follows. In Sec. II we present a basic  description of Dirac fermions and a brief introduction to the Floquet theory, emphasizing its application to this case.  In Sec. III we present the planar interface case, separating the cases of a wave impinging the interface normally or with a oblique angle. The formation of chiral currents at the interface is also discussed. In Sec. IV we present the anomalous Goos-H\"anchen shift that appears with electrons beams of a finite width. The results for the case of a circular irradiated spot are presented in Sec. V. Finally,  we summarize in Sec. VI. 

%%%%%%%%%%%%%%%%%%%%%%%%%%%%%%%%%%%%%%%%%%%%%%%%%%%%%%%%%%
\section{Low energy model: Driven Dirac Fermions}\label{section2}
%%%%%%%%%%%%%%%%%%%%%%%%%%%%%%%%%%%%%%%%%%%%%%%%%%%%%%%%%%
We consider a generic $2$D-system where the low energy excitations can be described by the following Hamiltonian
%%%%%%%%%%%%%%%%%%%%%%%%%%%%%%%%%%%%%%%%%%%%%%%%%%%%%%%%%%
\begin{equation}\label{1}
\hat{\cal{H}}=v_F\,\bm{\sigma}\cdot\bm{p}\,,  
\end{equation}
%%%%%%%%%%%%%%%%%%%%%%%%%%%%%%%%%%%%%%%%%%%%%%%%%%%%%%%%%%
where $\vf$ denotes the Fermi velocity, and  $\bm{\sigma}=(\sigma_x,\sigma_y)$ are Pauli matrices describing a pseudospin degree of freedom. For the sake of concreteness, we will take graphene as an example from hereon. In that case, the present model correspond to the low energy description of the carbon $p_z$-orbitals.% and $\vf\simeq 10^{6} m/s$.

We are interested in the case where there is a driving field described by the vector potential $\bm{A}(t)=\mathrm{Re}\left\{\bm{A}_0e^{\mathrm{i}\Omega t}\right\}$---for instance an electromagnetic field normally hitting  the graphene sheet. We assume a zero scalar potential. The electric field is then $\bm{E}(t){=}{-}(1/c)\partial_t \bm{A}(t)$ so that $E_0{=}|\bm{E}|{=}(\Omega/c) |\bm{A}_0|$. This is introduced in Eq.~\eqref{1} through the well known Peierls substitution
%%%%%%%%%%%%%%%%%%%%%%%%%%%%%%%%%%%%%%%%%%%%%%%%%%%%%%%%%%
\begin{equation}\label{4}
\bm{p} \rightarrow \bm{p} + \frac{e}{c}\,\bm{ A}(t)\,,
\end{equation}
%%%%%%%%%%%%%%%%%%%%%%%%%%%%%%%%%%%%%%%%%%%%%%%%%%%%%%%%%%
where $-e$ is the electron charge and $c$ the speed of light.  

We then have a Hamiltonian $\hat{\mathcal{H}}(t)$ that depends on time explicitly. In such a case, the energy of the system is no longer a conserved quantity and the usual approach of diagonalizing the Hamiltonian is no longer useful. Yet, for the special cases where the Hamiltonian is periodic in time one can apply the Floquet theory \cite{Shirley1965,Sambe1973} to reduce the calculation to an eigenvalue problem again. Just as a brief introduction we will present here the basic features of this method, a more extensive discussion can be found in Refs. \cite{Grifoni1998a,Kohler2005}.
%%%%%%%%%%%%%%%%%%%%%%%%%%%%%%%%%%%%%%%%%%%%%%%%%%%%%%%%%%
\begin{figure}[t]
\begin{center}
\includegraphics[width=0.9\columnwidth]{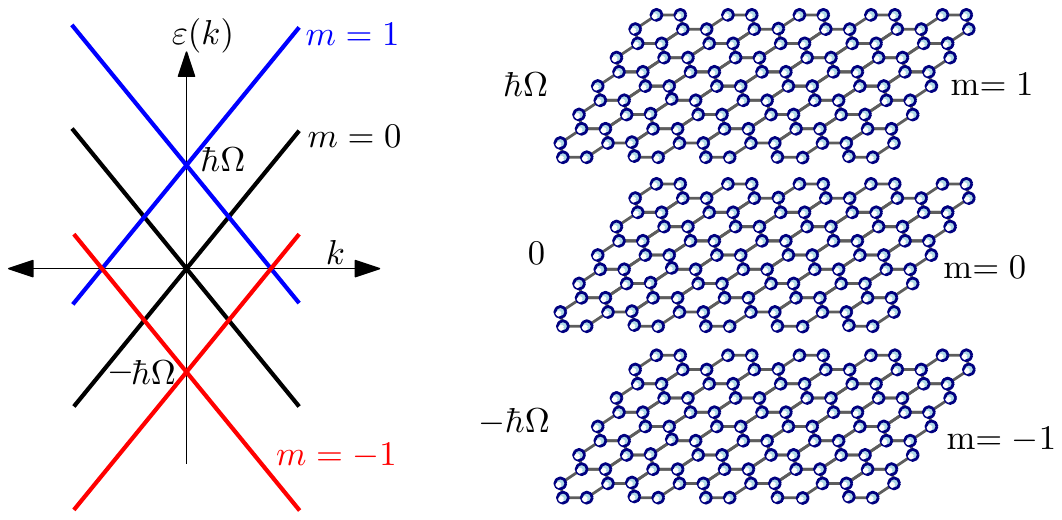}
\caption{(Color online) Visualization of the Floquet replicas as copies of the non irradiated system (graphene monolayer) coupled each other by means of absorption and emission of photons in the laser field \label{replicas}.}
\end{center}
\end{figure}
%%%%%%%%%%%%%%%%%%%%%%%%%%%%%%%%%%%%%%%%%%%%%%%%%%%%%%%%%%

Floquet theory  is a suitable approach for problems involving periodic time dependent  Hamiltonians $\hat{\mathcal{H}}(t){=}\hat{\mathcal{H}}(t{+}2\pi/\Omega)$, which can be written as a Fourier series in time $\hat{\mathcal{H}}(t){=}\sum_m \hat{\mathcal{H}}^{(m)}\,e^{i\,m\Omega t}$---here $m$ is an integer index. The solutions of the time dependent Schr{\"o}dinger equation $i\hbar\,\partial_t|\Psi\rangle{=}\hat{\mathcal{H}}(t)|\Psi\rangle$ can then be written as $|\Psi(t)\rangle=e^{-i\,\varepsilon t/\hbar}\,|\Phi(t)\rangle$, with $|\Phi(t)\rangle$ periodic in time with the same period as $\hat{\mathcal{H}}(t)$. The quantity $\varepsilon$ is called the quasienergy and, $|\Phi(t)\rangle$ satisfies the so-called Floquet equation
%%%%%%%%%%%%%%%%%%%%%%%%%%%%%%%%%%%%%%%%%%%%%%%%%%%%%%%%%%
\begin{equation}\label{2}
(\hat{\mathcal{H}}(t)-i\hbar\,\partial_t)|\Phi(t)\rangle=\varepsilon|\Phi(t)\rangle\,,
\end{equation}
%%%%%%%%%%%%%%%%%%%%%%%%%%%%%%%%%%%%%%%%%%%%%%%%%%%%%%%%%%
where the operator $\hat{\mathcal{H}}_F(t){=}\hat{\mathcal{H}}(t){-}i\hbar\,\partial_t $ is the  Floquet Hamiltonian.  Since $|\Phi(t)\rangle$ is periodic in time we can treat this time dependent problem as a time independent one by extending the Hilbert space and considering the product space $\cal{R}\otimes\cal{T}$ of the static (non driven) space $\cal{R}$ and the space $\cal{T}$ of functions periodic in time with period $T{=}2\pi/\Omega$. $\cal{T}$ can be spanned by the  basis functions $e^{i m\Omega t}$ with $m=0$, $\pm 1$, $\pm 2$, $\cdots$, 
while $\cal{R}$, on the other hand, is described by a set of kets $|\chi\rangle$, labeled by the quantum number $\chi$.  In our particular case, $\chi$ refers to the pseudospin degree of freedom (the sublattice  $A$ or $B$ in graphene). Hence $|\Phi(t)\rangle=\sum_{\chi,m} c_{\chi m} e^{i m\Omega t} |\chi\rangle$. 
By using the  Fourier series of $\hat{\mathcal{H}}(t)$ we can construct an explicit matrix representation for the Floquet Hamiltonian $\Ha_F$. In what follows we analyze the particular case of  Eq.~\eqref{1}.
Since we are mainly interested on the effects of a circularly polarized field we take the vector potential to be $\bm{A}(t)=A_0\,(\cos(\Omega t)\, \hat{\bm{x}}+\sin(\Omega t)\, \hat{\bm{y}})$. Hence, the time dependent Hamiltonian from Eq.~\eqref{1} can be written as
%%%%%%%%%%%%%%%%%%%%%%%%%%%%%%%%%%%%%%%%%%%%%%%%%%%%%%%%%%
\begin{equation}\label{5}
\hat{\mathcal{H}}(t)=\hat{\mathcal{H}}+\frac{e\vf A_0}{c}(\sigma_{\scriptscriptstyle{-}}\,e^{i\Omega t}+\sigma_{\scriptscriptstyle{+}}\,e^{-i\Omega t})\,,
\end{equation}
%%%%%%%%%%%%%%%%%%%%%%%%%%%%%%%%%%%%%%%%%%%%%%%%%%%%%%%%%%
with $\sigma_\pm=(\sigma_x{\pm}i\sigma_y)/2$. Expanding the exponentials and passing to the direct product basis $|\chi,m\rangle$, we get the following matrix representation
%%%%%%%%%%%%%%%%%%%%%%%%%%%%%%%%%%%%%%%%%%%%%%%%%%%%%%%%%%
\begin{equation}\label{6}
\hat{\mathcal{H}}_F=\left[
\begin{array}{ccccc}
\ddots & \vdots          & \vdots  & \vdots          & \reflectbox{$\ddots$} \\
\dots  & \displaystyle{\hat{\mathcal{H}}{+}\hbar\Omega} & \eta\hbar\Omega\,\sigma_{\scriptscriptstyle{-}}     & 0            & \dots \\
\dots  & \eta\hbar\Omega\,\sigma_{\scriptscriptstyle{+}}         & \hat{\mathcal{H}}     & \eta\hbar\Omega\,\sigma_{\scriptscriptstyle{-}}            & \dots \\
\dots  & 0          & \eta\hbar\Omega\,\sigma_{\scriptscriptstyle{+}}  & \hat{\mathcal{H}}{-}\hbar\Omega & \dots \\
\reflectbox{$\ddots$} & \vdots          & \vdots  & \vdots          & \ddots
\end{array}
\right]\,.
\end{equation}
%%%%%%%%%%%%%%%%%%%%%%%%%%%%%%%%%%%%%%%%%%%%%%%%%%%%%%%%%%
Here we have defined the dimensionless parameter $\eta{=}e\vf A_0/c\hbar\Omega$ to quantify the strength of the laser field.
%%%%%%%%%%%%%%%%%%%%%%%%%%%%%%%%%%%%%%%%%%%%%%%%%%%%%%%%%%
\begin{figure}[t]
\begin{center}
\includegraphics[width=0.65\columnwidth]{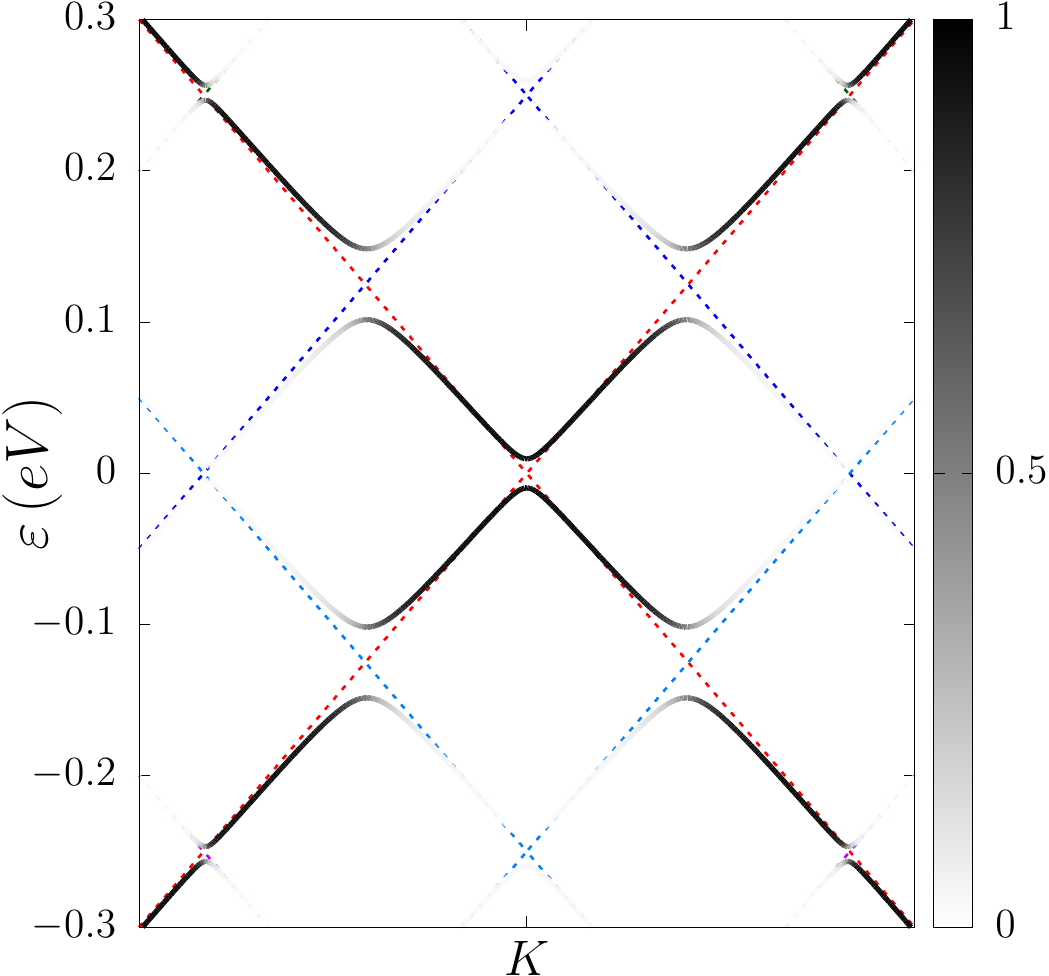}
\caption{Spectral density projected onto the $m=0$ replica for irradiated graphene (solid line), and uncoupled Floquet quasienergy bands (color dashed lines). Dynamical gaps appear at quasienergies where the Floquet replicas become degenerate  \label{gaps_Floquet}.}
\end{center}
\end{figure}
%%%%%%%%%%%%%%%%%%%%%%%%%%%%%%%%%%%%%%%%%%%%%%%%%%%%%%%%%%
$\Ha_F$ is then equivalent to a static Hamiltonian formed by copies (Floquet replicas) of the original system shifted in energy by multiples of $\hbar\Omega$ and coupled between them to nearest neighbors  through the parameter $\eta$ (see Fig.~\ref{replicas}). The spectrum of the uncoupled set of Floquet replicas crosses each other at specific quasienergies, making them degenerate. The driving field lifts these degeneracies and therefore the corresponding Floquet spectrum (for $\eta\neq0$) presents gaps at specific quasienergies~\cite{Usaj2014a}. This is shown in Fig.~\ref{gaps_Floquet} where replicas  from $m=-2$ through $m=2$ have been considered. 

In this work, we will focus on the gap of order $\eta \hbar\Omega$ that appears at $\varepsilon{\sim}\hbar\Omega/2$ where the replicas $m=0$ and $m=1$ cross each other. In that case, since we will only consider the limit $\eta\ll1$, it is sufficient to restrict the Floquet Hamiltonian to those replicas ($m=0$ and $m=1$), which will allows us to obtain  analytical results. Therefore, we are left with the following  reduced Floquet matrix
%%%%%%%%%%%%%%%%%%%%%%%%%%%%%%%%%%%%%%%%%%%%%%%%%%%%%%%%%%
\begin{equation}\label{7}
\tilde{\mathcal{H}}_F=\left(
\begin{array}{cccc}
\hbar\Omega & \vf \pmenos & 0 & 0\\
\vf \pmas &\hbar\Omega &  \eta\hbar\Omega & 0 \\
0& \eta\hbar\Omega &0& \vf \pmenos \\
0&0 & \vf \pmas & 0\end{array}
\right)\,,
\end{equation}
%%%%%%%%%%%%%%%%%%%%%%%%%%%%%%%%%%%%%%%%%%%%%%%%%%%%%%%%%%
with $p_{\pm}=p_x\pm i p_y=-i \hbar(\partial_x\pm i \partial_y)$, that corresponds to a counterclockwise circular polarization. Once the wavefunction for this case is obtained, one can get the corresponding one for the clockwise polarization by swapping the spinors components on each channel and making the substitution $p_y{\rightarrow}-p_y$.

%%%%%%%%%%%%%%%%%%%%%%%%%%%%%%%%%%%%%%%%%%%%%%%%%%%%%%%%%%
\begin{figure}[t]
\begin{center}
\includegraphics[width=0.95\columnwidth]{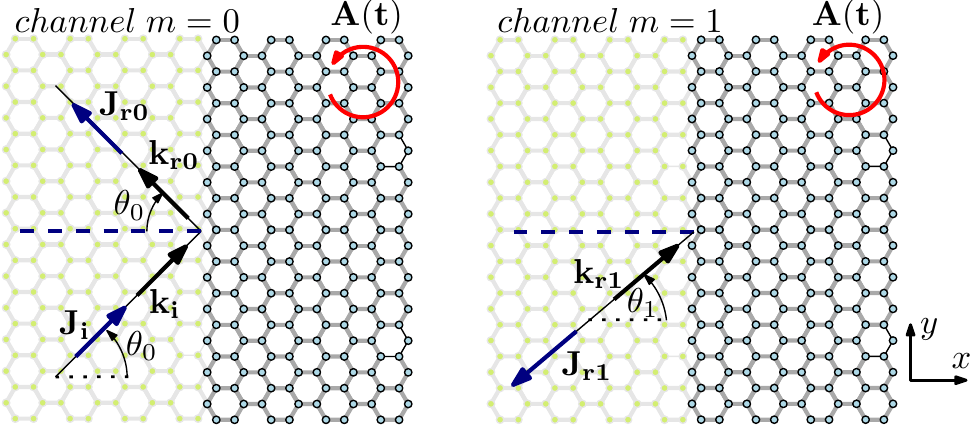}
\caption{Geometry of the scattering problem  with oblique incidence. The darker region corresponds to the irradiated area. Due to the conservation $k_y$, the reflected wave in the $m=1$ channel  follows a path similar to that of the incident wave. In particular, $\theta_1=\theta_0$ when $\mu=0$. \label{fig1}}
 \label{geometry-plane}
\end{center}
\end{figure}
%%%%%%%%%%%%%%%%%%%%%%%%%%%%%%%%%%%%%%%%%%%%%%%%%%%%%%%%%%

The solutions of the eigenvalue equation (cf. Eq.~\eqref{2}) with  Hamiltonian~\eqref{7} depend on the geometry of the problem  under consideration. In the following sections we will find such solutions, with asymptotic incoming states on the non-irradiated region, for two cases: (i) When the interface between irradiated and non irradiated regions is a straight line, and (ii) when the irradiated region is circular. 

%%%%%%%%%%%%%%%%%%%%%%%%%%%%%%%%%%%%%%%%%%%%%%%%%%%%%%%%%%
\section{Straight interface}
%%%%%%%%%%%%%%%%%%%%%%%%%%%%%%%%%%%%%%%%%%%%%%%%%%%%%%%%%%
We start by analyzing the  simplest situation, which consists of an infinite  sheet of, say,  graphene ($xy$ plane), with the half-plane $x{>}0$  irradiated by a laser field described by the vector potential $\bm{A}(t)=A_0\,(\cos(\Omega t)\, \hat{\bm{x}}+\sin(\Omega t)\, \hat{\bm{y}})$, as shown in Fig.~\ref{geometry-plane}. Within the approximation where only the $m=0$ and $m=1$ Floquet channels are retained, the problem reduces to find a four component Floquet wave function.  Since we are interested in describing the scattering states, and to simplify the problem, we will look for solutions where the asymptotic incident states belong only to the  $m=0$ channel. Hence, our main goal is to find the probability of scattering in each channel after the dispersion by the interface created by the inhomogeneous laser field.   
To solve the problem we then need to resolve the eigenvalue equation with $\tilde{\mathcal{H}}_F$ defined in Eq.~\eqref{7} in both regions, $x{>}0$ ($A_0{\neq} 0$) and $x{<}0$ ($A_0=0$), and then match these solutions at the boundary $x=0$.

For clarity, we will discuss separately the cases of normal and oblique incidence as the former is mathematically simpler and will give us a good insight of the physics involved.

%%%%%%%%%%%%%%%%%%%%%%%%%%%%%%%%%%%%%%%%%%%%%%%%%%%%%%%%%%
\subsection{Normal Incidence}
%%%%%%%%%%%%%%%%%%%%%%%%%%%%%%%%%%%%%%%%%%%%%%%%%%%%%%%%%%
In this case we have that the $k_y$ component of wavevector is zero so that the problem becomes essentially one dimensional.  Let us introduce the dimensionless parameter $\mu$ through the relation $\varepsilon=\hbar\Omega(1+\mu)/2$ . 
In the non irradiated region ($x{<}0$) there is no coupling between channels ($A_0=0$) so that they can be solved independently. Since we consider the incident particle  to be in the $m=0$ channel, we take the incident component of the $m=1$ channel equal to zero.Then, the $x<0$ solution for the $m-$replica can be obtained from the following matrix equation
%%%%%%%%%%%%%%%%%%%%%%%%%%%%%%%%%%%
\begin{equation}
\left(
  \begin{array}{cc}
      m\hbar\Omega & \vf\pmenos \\
      \vf \pmas     & m\hbar\Omega
  \end{array}    
   \right)
%  \binom{u_A^{(m)}}{ u_B^{(m)}}
\Phi_m^<
   =\varepsilon 
%  \binom{u_A^{(m)}}{ u_B^{(m)}}\,,
\,\Phi_m^<\,.
\end{equation}
%%%%%%%%%%%%%%%%%%%%%%%%%%%%%
In our case  these solutions read
%%%%%%%%%%%%%%%%%%%%%%%%%%%%%%%%%%
\begin{eqnarray}\label{sol-no-rad}
  \Phi_1^<(x) &=&r_1\, \binom{1}{-1} e^{i(\hbar\Omega-\varepsilon)x/\hbar \vf} , \nonumber\\
  \Phi_0^<(x) &=&\binom{1}{1}e^{i\varepsilon x/\hbar \vf}+r_0\, \binom{1}{-1} e^{-i\varepsilon x/\hbar \vf}\,.
\end{eqnarray}
%%%%%%%%%%%%%%%%%%%%%%%%%%%%%%%%%%%%%%%%%%%%%%%%%%%%%%%%%%
Here a global normalization factor has been omitted. The complex quantities $r_0$ and $r_1$ are the reflection amplitudes in each channel, the corresponding reflection coefficients being $R_0=|r_0|^2$ and $R_1=|r_1|^2$. Here we stress that the flux direction of the particles on each channel  must be determined by calculating the current $J_{x}^{(m)}=\Phi_m^\dagger \sigma_x\Phi_m$.   

In the region $x{>}0$ one needs to solve the complete Floquet equation $\tilde{\mathcal{H}}_F \psi=\varepsilon \psi$, $\psi=(\Phi_1^>,\Phi_0^>)$, coupling both channels. This leads to four independent solutions along with four integration constants $C_i$,
\begin{widetext}
%%%%%%%%%%%%%%%%%%%%%%%%%%%%%%%%%%%%%%%%%%%%%%%%%%%%%%%%%%
\begin{eqnarray}
\label{rock}
\nonumber
  \Phi_1^>(x)&=&\, C_1 \, \dbinom{-\frac{k_+}{k_0^-}}{1}e^{ik_+x} + 
  C_2\, \dbinom{\frac{k_+}{k_0^-}}{1} e^{-ik_+x} +
  C_3\, \dbinom{-\frac{k_-}{k_0^-}}{1} e^{ik_-x} + 
  C_4\, \dbinom{\frac{k_-}{k_0^-}}{1} e^{-ik_-x}\, ,  \\
    \Phi_0^>(x)&=&\, C_1\Pi_+ \, \dbinom{1}{\frac{k_+}{k_0^+}} e^{ik_+x}+ 
  C_2\Pi_+\, \dbinom{1}{-\frac{k_+}{k_0^+}} e^{-ik_+x}+ 
  C_3\Pi_-\, \dbinom{1}{\frac{k_-}{k_0^+}} e^{ik_-x} + 
  C_4\Pi_-\, \dbinom{1}{-\frac{k_-}{k_0^+}} e^{-ik_-x}\, ,
\end{eqnarray}
%%%%%%%%%%%%%%%%%%%%%%%%%%%%%%%%%%%%%%%%%%%%%%%%%%%%%%%%%%
\end{widetext}
with $k_0=\Omega/2\vf$, $k_0^\pm=k_0(1\pm \mu)$ and 
%%%%%%%%%%%%%%%%%%%%%%%%%%%%%%%%%%
\bea\label{equs}
\Pi_\pm &=& \frac{\mu\pm\sqrt{\mu^2(1+\eta^2)-\eta^2}}{\eta(1-\mu)},\\
 k_\pm  &=& k_0\sqrt{1{+}\mu^2 \pm 2\,\sqrt{-\eta^2+\mu^2(1+\eta^2)}}\,.
\eea
%%%%%%%%%%%%%%%%%%%%%%%%%%%%%%%%
The way in which the square roots are taken depends on the particular values of $\mu$ and $\eta$ as we will see later.
The  continuity of the wave function at $x=0$ results in a system of {\it four} equations from where $r_0$, $r_1$, and the four coefficients $C_i$ ({\it six} unknowns) need to be determined. The number of equations is clearly not sufficient and one needs to impose some additional requirements to eliminate two integration constants.  These requirements are fairly obvious and can be better described by analyzing separately the cases where the incident particle has a quasienergy that falls inside or outside the dynamical gap. The former situation can be written in terms of $\mu$ as
%%%%%%%%%%%%%%%%%%%%%%%%%%%%%%%
\be
|\mu|\leq\frac{\eta}{\sqrt{1+\eta^2}}\,.
\ee
%%%%%%%%%%%%%%%%%%%%%%%%%%%%%%%
%%%%%%%%%%%%%%%%%%%%%%%%%%%%%%%%%%%%%%%%%%%%%%%%%%%%%%%%%%
\subsubsection*{Inside the Dynamical Gap}
%%%%%%%%%%%%%%%%%%%%%%%%%%%%%%%%%%%%%%%%%%%%%%%%%%%%%%%%%%
In this case,  there are no propagating states in the irradiated region and so we require that the wavefunction  vanishes asymptotically as $x \rightarrow \infty$. This requires the solutions in Eqs.~\eqref{equs} to be complex with $k_+$ and $k_-$ complex conjugate of each other. The only two well behaved solutions for $x{\rightarrow} \infty$ are those with the factors $e^{ik_+x}$ and $e^{-ik_-x}$, with $k_+=k_0\left(1{+}\eta^2{+}2i\sqrt{\eta^2{-}\mu^2(1{+}\eta^2)}\right)^{1/2}$ and $k_-=k_+^*$ (the complex square root is taken in the principal branch). On the contrary, the remaining solutions must be discarded  since they diverge in that limit (we take $C_2=C_3=0$).  These solutions  give no total current  flowing into the irradiated region (both channels combined) and therefore the particles are fully backscattered with a fraction $R_0$ and $R_1$ in the corresponding channels, $R_0+R_1=1$.
%%%%%%%%%%%%%%%%%%%%%%%%%%%%%%%%%%%%%%%%%%%%%%%%%%%%%%%%%%
\subsubsection*{Outside the Dynamical Gap}%%%%%%%%%%%%%%%%%
%%%%%%%%%%%%%%%%%%%%%%%%%%%%%%%%%%%%%%%%%%%%%%%%%%%%%%%%%%
When the quasienergy of the incident particle lies outside the dynamical gap we expect propagation in the irradiated region and therefore a non zero probability current for $x{>}0$. If $\mu$ is not so large (otherwise the two-replica approximation in not valid), $k_1$ and $k_2$ are both real and can be taken positive
%%%%%%%%%%%%%%%%%%%%%%%%%%%%%%%
\begin{equation}
k_{\pm} =k_0\,\sqrt{1+\mu^2\pm2\sqrt{\mu^2(1+\eta^2)-\eta^2}}\,.  
%k_2 &=&k_0\,\sqrt{1+\mu^2-2\sqrt{\mu^2(1+\eta^2)-\eta^2}}\,.
\end{equation}
%%%%%%%%%%%%%%%%%%%%%%%%%%%%%%%%
After matching solutions with Eqs.~\eqref{sol-no-rad} at $x=0$, we have to discard variables in order to have a consistent system of equations.
The physical requirement here is that the total current in the irradiated region, this is the time-averaged current coming from the complete wave function $\Phi(t)=(\Phi_0^>+e^{i\Omega t}\,\Phi_1^>)\,e^{i\varepsilon t\hbar}$, be directed towards the positive $\hat{\bm{x}}$ direction and with an increasing proportion in channel $m=0$ as $\eta$ vanishes. This latter condition ensures that as the electromagnetic field fades away the electron beam, which is coming in the  $m=0$ channel, is fully transmitted in the same channel. This of course tells us that the reflection coefficients $r_0$ and $r_1$ also vanish. As a final remark, this total current must arise in a continuous way from its zero value inside the gap and increase as we move away the gap (of course keeping $\mu$ small). This transmitted current gives rise to a transmittance $T$ and the conservation of flux probability leads to $  R_0+R_1+T=1$.

\begin{figure}[t]
\includegraphics[width=0.9\columnwidth]{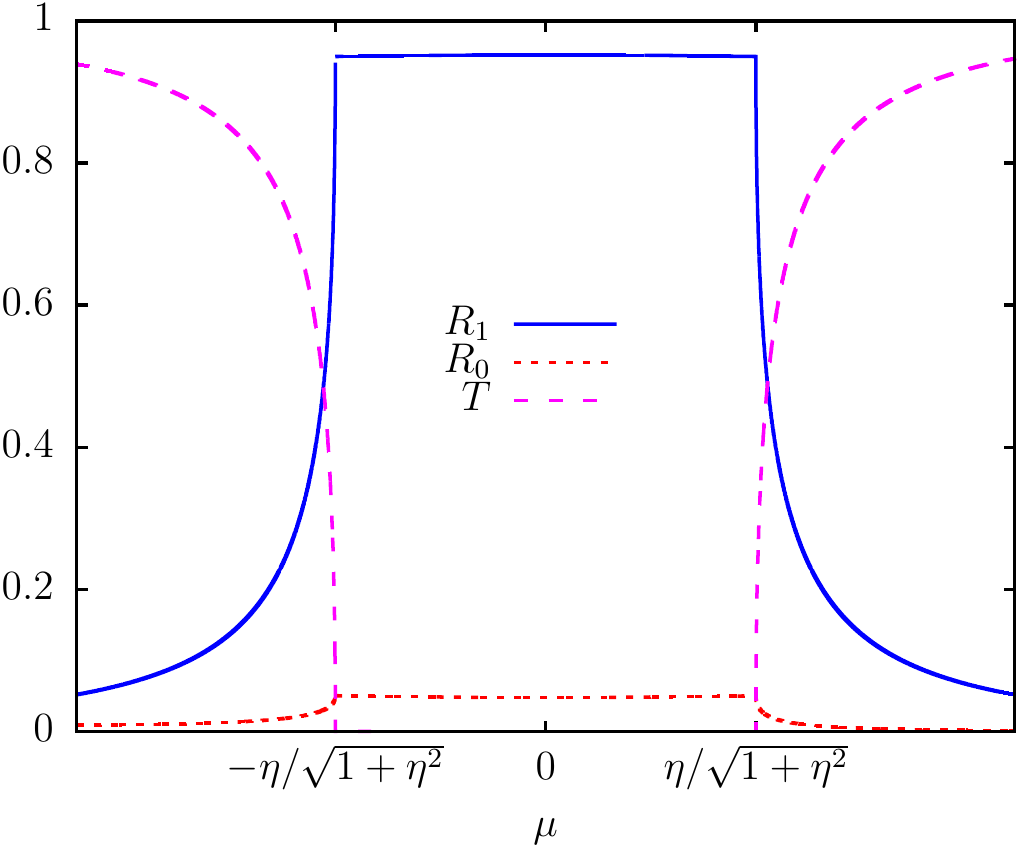}        \\
\caption{Reflection and transmission coefficients in channels $m=1$ and $m=0$ as a function of the dimensionless parameter $\mu$ given by $\varepsilon=\hbar\Omega(1{+}\mu)/2$. The range $| \mu| \leq \eta/\sqrt{1+\eta^2}$ defines the dynamical gap.  It must be noted the lack of symmetry around $\mu=0$, more clearly in $R_0$ \label{reflectances}.}
\end{figure}

The condition of a current towards the $\hat{\bm{x}}$  direction permits to reduce the number of unknowns $C_i$ ($i=1...4$). Using the equation $J_x=\Phi^\dagger \sigma_x\Phi$, the transmitted current in the irradiated region can be written as the sum of two terms
%%%%%%%%%%%%%%%%%%%%%%%%%%%%%%%%%%%%%%%%%%%%%%%%%%%%%%%%%%
\begin{equation}
J_x^T=J_x^{(1)}+J_x^{(2)},
\end{equation}
%%%%%%%%%%%%%%%%%%%%%%%%%%%%%%%%%%%%%%%%%%%%%%%%%%%%%%%%%%
where
%%%%%%%%%%%%%%%%%%%%%%%%%%%%%%%%%%%%%%%%%%%%%%%%%%%%%%%%%%
\begin{eqnarray}
J_x^{(1)} &=&\frac{2}{1{-}\mu}\left[ k_-|C_4|^2\left(1{-}\frac{\Pi_-}{\Pi_+}\right) - k_+|C_1|^2\left(1{-}\frac{\Pi_+}{\Pi_-}\right) \right]\,, \notag\\
J_x^{(2)} &=&\frac{2}{1{-}\mu}\left[ k_+|C_2|^2\left(1{-}\frac{\Pi_+}{\Pi_-}\right) - k_-|C_3|^2\left(1{-}\frac{\Pi_-}{\Pi_+}\right) \right]\,. \notag\\
\end{eqnarray}
%%%%%%%%%%%%%%%%%%%%%%%%%%%%%%%%%%%%%%%%%%%%%%%%%%%%%%%%%%
These currents have the property that when $\mu{>}\eta/\sqrt{1+\eta^2}$ (above the gap), $J_x^{(1)}{>}0$ and $J_x^{(2)}{<}0$, whereas when $\mu{<}{-}\eta/\sqrt{1+\eta^2}$ (below the gap), $J_x^{(1)}{<}0$ and $J_x^{(2)}{>}0$. So, in order to have a positive current (in the  $\hat{\bm{x}}$ direction), we have to make $C_2=C_3=0$ when we are above the gap and $C_1=C_4=0$ when we are below it.  With these considerations we can calculate $r_0$, $r_1$ and the relevant $C_i$. The transmittance, as usual, is calculated as $T=|J_x^T/J_{\mathrm{inc}}|$.
In Fig.~\ref{reflectances} we plot the reflection coefficients in both channels, $R_0$ and $R_1$, and the transmittance $T$ inside and outside the dynamical gap (indicated by the limits $\pm\eta/\sqrt{1+\eta^2}$ in the $\mu$ axis). 
Inside the gap, $R_1$ greatly exceeds $R_0$. That is, particles coming from channel $m=0$ are mostly backscattered through the channel $m=1$. The origin of this is both the conservation of $k_y$ and the pseudospin flip induced by the electromagnetic field~\cite{Katsnelson2006}.
As we move apart from the gap, $R_0$ and $R_1$ no longer add up to $1$ since in that case there are currents flowing into the irradiated region, so that $T\neq0$---this transmittance is no longer defined for each channels due to the coupling imposed by the electromagnetic field. It is worth noting that $T$ increases continuously from zero precisely at the border of the gap, as  expected. Moreover, the non analytic behavior of the reflection coefficients at the gap border resembles that found for the stationary problem of the scattering of particles from a potential step  when the energy of the incoming particles  overcomes the potential height.

%%%%%%%%%%%%%%%%%%%%%%%%%%%%%%%%%%%%%%%%%%%%%%%%%%%%%%%%%%%
%%%%%%%%%%%%%%%%%%%%

%%%%%%%%%%%%%%%%%%%%%%%%%%%%%%%%
\subsection{Oblique Incidence}%%
%%%%%%%%%%%%%%%%%%%%%%%%%%%%%%%%
Let us now consider the case when the incident particle hits the interface at a given angle $\theta_0$ (measured counterclockwise from the horizontal direction, see Fig.~\ref{geometry-plane}) and let us first solve the uncoupled Floquet equation for channels $m=0$ and $m=1$. As before, we will use a spinor plane wave of the form $\Phi_m(x,y)=\phi_m\,e^{i(k_x^{(m)}x+k_yy)}$. For a given quasi energy $\varepsilon$ the components $k_x^{(0)}$ and $k_x^{(1)}$ for each channel satisfy
%%%%%%%%%%%%%%%%%%%
\begin{eqnarray}
{k_x^{(0)}}^2+k_y^2 &=&(k_0^{+})^2 \\
{k_x^{(1)}}^2+k_y^2 &=&(k_0^{-})^2\,,
\end{eqnarray}
%%%%%%%%%%%%%%%%%%
and writing $k_y=k_0^+\sin\theta_0$ we get 
%%%%%%%%%%%%%%%%%%%%%
\begin{eqnarray}
k_x^{(0)} &=& k_0^+\cos\theta_0 \\
k_x^{(1)} &=& \sqrt{(k_0^{-})^2-(k_0^{+})^2\sin^2\theta_0}\,.
\end{eqnarray}
From the last equation is it clear that in order to have a real value of $k_x^{(1)}$ (and so a propagating wave in the  $m=1$ channel), the incident angle $\theta_0$ must satisfy
%%%%%%%%%%%%%%%%%%%%%%
\begin{equation}\label{cond}
  |\sin\theta_0|\leq \frac{1-\mu}{1+\mu}\,.
\end{equation}
%%%%%%%%%%%%%%%%%%%%%%
%%%%%%%%%%%%%%%%%%%%
\begin{figure*}[t]
\begin{center}
\includegraphics[width=0.9\textwidth]{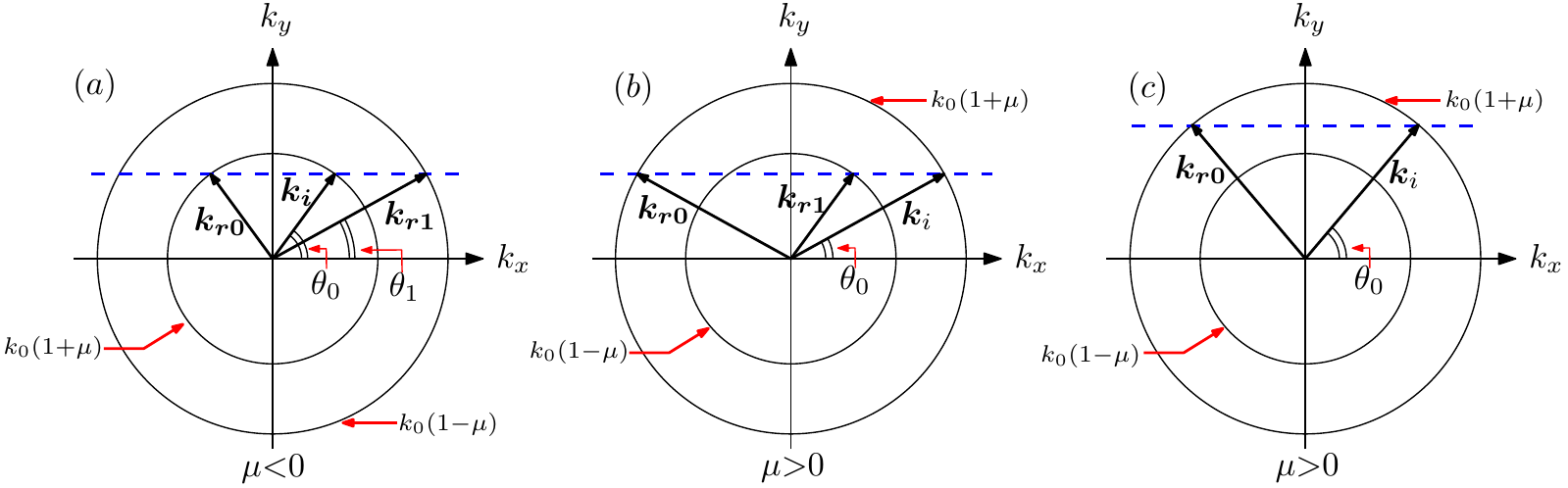}
\caption{Curves of constant quasi energy for: (a) $\mu{<}0$ (below the center of the gap) and (b) $\mu{>}0$ (above the center of the gap). In (a) the value of $\bm{k_{r1}}$ is always well defined for all possible values of $\bm{k_i}$, meaning that the reflected wave in the channel $m=1$ is a traveling one. In (b) there is a critical angles $\sin\theta_c=(1{-}\mu)/(1{+}\mu)$ above which $\bm{k}_{r1}$ is ill defined. Physically, this means that  $\bm{k}_{r1}$ is complex  and  the reflected wavefunction in channel $m=1$ is evanescent.  \label{level_diagram}}
\end{center}
\end{figure*}
%%%%%%%%%%%%%%%%%%%%%%%%%%%%%%
It can be seen that this requirement is trivially fulfilled for all values $-\pi/2\leq\theta_0\leq\pi/2$ when $\mu<0$ (below the center of the gap). However, when $\mu{>}0$ (above the center of the gap), there is a critical angle $\theta_c$ satisfying
%%%%%%%%%%%%%%%%%%%%%%
\begin{equation}\label{theta_c}
\sin\theta_c=\frac{1{-}\mu}{1{+}\mu}\,,
\end{equation}
%%%%%%%%%%%%%%%%%%%%%%
so that for $|\theta_0|\leq\theta_c$ the solution $k_x^{(1)}$ is real and the reflected wave in the  $m=1$ channel a traveling wave, while for $|\theta_0|>\theta_c$, $k_x^{(1)}$ is pure imaginary and then we get an evanescent solution that must be chosen in such a way that it goes to zero as $x{\rightarrow} -\infty$. This gives a boundary wave in the channel $m=1$ that propagates along the interface. 
The solutions in the non illuminated region can be readily obtained 
%%%%%%%%%%%%%%%%%%%%%%%%%%%%%%%%%%
\bea\label{zero}
\Phi_1^{<}(x) &=& r_1 \binom{-e^{-i\theta_1/2}}{e^{i\theta_1/2}} e^{ik_x^{(1)}x}\,,\\
\nonumber
\Phi_0^<(x)  &=&\binom{e^{-i\theta_0/2}}{e^{i\theta_0/2}} e^{ik_{x}^{(0)}x} {+} r_0 \binom{-e^{i\theta_0/2}}{e^{-i\theta_0/2}} e^{-ik_{x}^{(0)}x}\,,
\eea
%%%%%%%%%%%%%%%%%%%%%%%%%%%%%%%%%%%%%%%%
where for the sake of simplicity we omitted an overall factor $e^{ik_yy}$.
Here, $r_0$ and $r_1$ determine again the reflection coefficients. The form of $\Phi_1^{<}(x)$ ensures that there is no incident wave in the $m=1$ channel. Moreover, when $\theta_0=0$, this solution recovers that in Eqs.~\eqref{sol-no-rad}, as it should.
In the above expressions, we have introduced an extra angle $\theta_1$ through the relation $k_y=k_0^-\sin\theta_1$, which gives $\sin\theta_1=\frac{(1{+}\mu)}{(1{-}\mu)}\sin\theta_0$, and defines  $k_x^{(1)}=k_0^-\cos\theta_1$. 
Clearly, $\theta_1$ is always real when $\mu{\leq} 0$ (in particular $\theta_1=\theta_0$ when $\mu=0$) and when both $\mu{>}0$ and $|\theta_0|\leq\theta_c$. Otherwise $\theta_1$ is complex and to obtain the appropriate asymptotic behavior we choose the branch 
%%%%%%%%%%%%%%%%%%%%%%%
\begin{equation}
\cos\theta_1=-i\sqrt{\frac{\sin^2\theta_0}{\sin^2\theta_c}-1}\,.
\end{equation}
%%%%%%%%%%%%%%%
A graphical view to arrive at the same conclusions is depicted in Fig.~\ref{level_diagram}, where quasienergy surface of the Floquet replicas at a certain value of $\varepsilon$ are shown.  Figure~\ref{level_diagram}(a) show these surfaces when $\mu{<}0$ (equivalently $\varepsilon{<}\hbar\Omega/2$). In this case the inner (outer) circle corresponds to the replica $m=0$ ($m=1$). From its  definition $|k_y|\leq k_0^+$ in the whole range of incident angles ($|\theta_0|\leq\pi/2$). This means that for any given value of $k_y$ (the horizontal dotted blue line) one can always find the corresponding wave vectors of the reflected waves $\bm{k}_{r0}$ and $\bm{k}_{r1}$, meaning that they are both real and the solution is thereby a traveling wave.
When $\mu{>}0$ the situation is reversed, and the quasienergy surface of channel $m=1$ lies inside the one corresponding to channel $m=0$, as shown in Fig.~\ref{level_diagram}(b). It is apparent that if $|k_y|\leq k_0^-$, it is still possible to find real solutions for both $\bm{k}_{r0}$ and $\bm{k}_{r1}$, as before. However, if $|k_y|{>} k_0^-$ (Fig.~\ref{level_diagram}(c)) there is no real solution for $\bm{k}_{r1}$. This in turns implies that for $|\theta_0|>\theta_c$ we have $R_1=0$.

On the other hand, the most general solution of the Floquet equation in the irradiated region   can be written as
\begin{widetext}
%%%%%%%%%%%%%%%%%%%%%%%%%%%%%%%%%%%%%%%%%%%%%%%%%%%%%%%%%%
\begin{eqnarray}\label{inside}
\Phi_1^{>}(x) &=&C_{1}e^{ik_+x} \left(\begin{array}{c}
  \frac{(i k_y-k_+)}{k_0^-} \\
  1
\end{array} \right) + C_{2} e^{-ik_+x} \left(\begin{array}{c}
\frac{ik_y+k_+}{k_0^-} \\
  1 
\end{array} \right) +  
C_{3} e^{ik_-x} \left(\begin{array}{c}
  \frac{ik_y-k_-}{k_0^-} \\
  1
\end{array} \right) + C_4e^{-ik_-x} \left(\begin{array}{c}
\frac{ik_y+k_-}{k_0^-} \\
  1 
\end{array} \right)  \,,  \\
\nonumber 
\Phi_0^{>}(x) &=&\Pi_+ C_{1}e^{ik_+x} \left(\begin{array}{c}
   1 \\
  \frac{(i k_y+k_+)}{k_0^+} 
\end{array} \right) + \Pi_+ C_{2} e^{-ik_+x} \left(\begin{array}{c}
  1\\
  \frac{ik_y-k_+}{k_0^+} 
\end{array} \right) + 
\Pi_- C_{3} e^{ik_-x} \left(\begin{array}{c}
  1 \\
  \frac{ik_y+k_-}{k_0^+} 
\end{array} \right)  + \Pi_- C_4e^{-ik_-x} \left(\begin{array}{c}
1 \\
\frac{ik_y-k_-}{k_0^+} 
\end{array} \right)  \,.
\end{eqnarray}
%%%%%%%%%%%%%%%%%%%%%%%%%%%%%%%%%%%%%%%%%%%%%%%%%%%%%%%%%%
\end{widetext}
Here, for simplicity, we will restrict ourselves to the case where the quasi energy $\varepsilon$ lies inside the dynamical gap. Then we have
%%%%%%%%%%%%%%%%%%%%%%%%%%%%%%%%%%%%%%%%%%%%%%%%%%%%%%%%%%
\begin{figure}[t]
\begin{center}
\includegraphics[width=0.85\columnwidth]{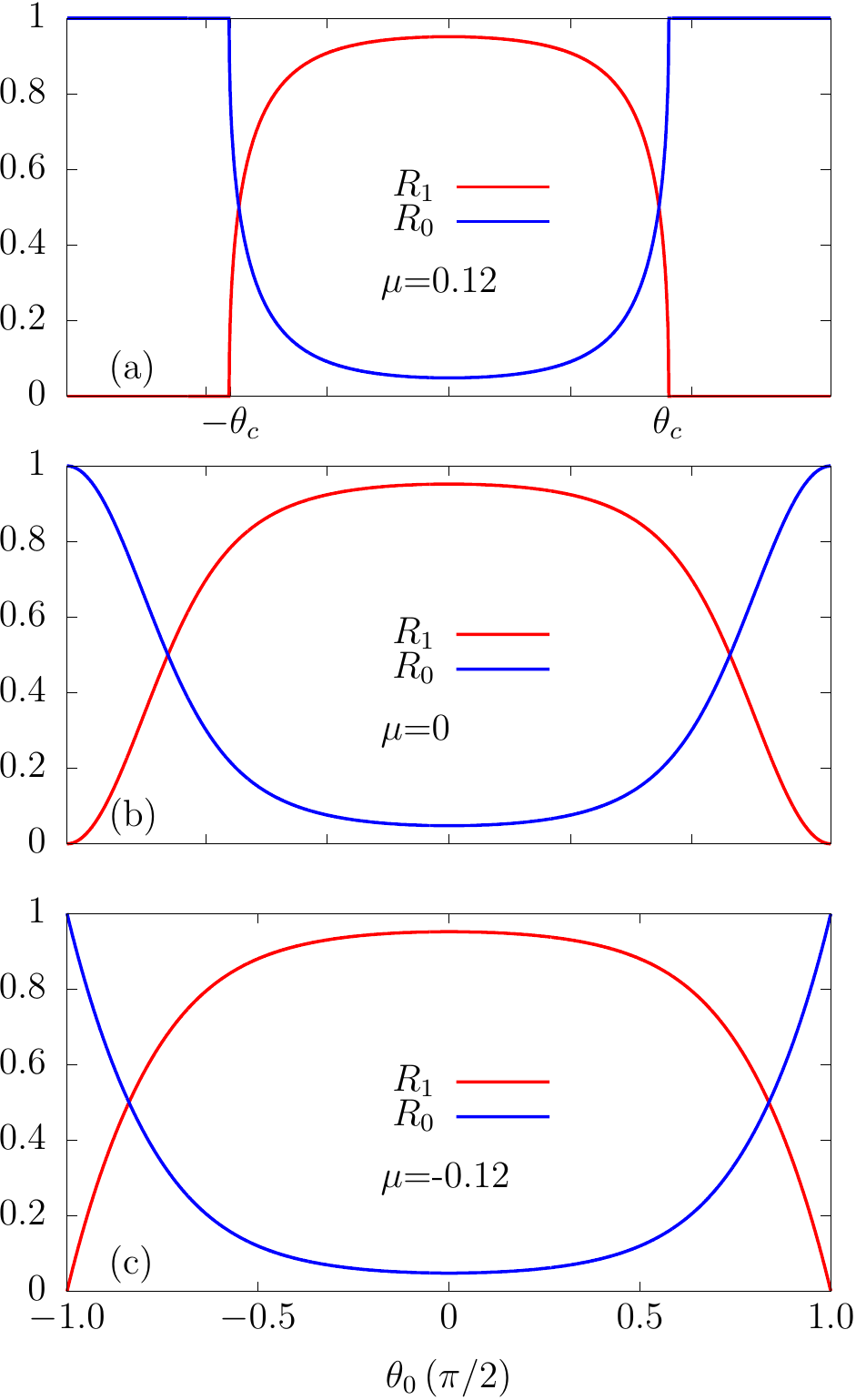}
\caption{Reflectances  in channels  $m=0$ and $m=1$ as a function of $\theta_0$ for three different values of $\mu$. (a) When $\mu{>}0$ there is an critical angle $\theta_c$ beyond which the reflectance in channel $m=1$ vanishes. This corresponds to the transition of the wave function in channel $m=1$ from a traveling wave to an evanescent one. (b) If $\mu=0$ we get the same result as in normal incidence for $\theta_0=0$, as expected, although when $\theta_0$ gets closer to $\pm\pi/2$ the opposite result is obtained. (c) The same as (b) but for $\mu{<}0$. In all cases we used $\eta=\mbox{0.23}$ and thus the limit of the dynamical gap is $\eta/\sqrt{1+\eta^2}\approx 0.22$ \label{R_T}.}
\end{center}
\end{figure}
%%%%%%%%%%%%%%%%%%%%%%%%%%%%%%%%%%%%%%%%%%%%%%%%%%%%%%%
%%%%%%%%%%%%%%%%%%
\begin{eqnarray}
k_+   &=& k_0 \sqrt{-\frac{k_y^2}{k_0^2}+1+\mu^2+2i\,\sqrt{\eta^2{-}\mu^2(1+\eta^2)}}, \\
\Pi_\pm &=& \frac{\eta\,(1{+}\mu)}{\mu\mp i\sqrt{\eta^2{-}\mu^2(1+\eta^2)}}\,,
\end{eqnarray}
%%%%%%%%%%%%%
with $k_-= k_+^*$. 
As before, we reduce the number of constants with some physical requirements. Since we have restricted ourselves to values of $\mu$ inside the dynamical gap,  the physical solution vanish as $x\rightarrow +\infty$, that implies $C_2=C_3=0$.

In Fig.~\ref{R_T} we see plots of reflectances $R_0$ and $R_1$  for the cases $\mu=0$ and $\mu=\pm0.12$ taking $\eta=0.23$. As we pointed out before, when $\mu{>} 0$ we find that there is a total reflection on the $m=0$ channel for $|\theta_0|>\theta_c\sim 0.3\pi$.  At normal incidence ($\theta_0=0$) we recover the known result that the electrons are mainly back scattered in channel $m=1$ in all three cases. 
%%%%%%%%%%%%%%%%%%%%%%%%%%%%%%%%%%%%%%%%%%%%%
\subsection{Chiral currents along the interface}
\begin{figure}[t]
\begin{center}
\includegraphics[width=0.9\columnwidth]{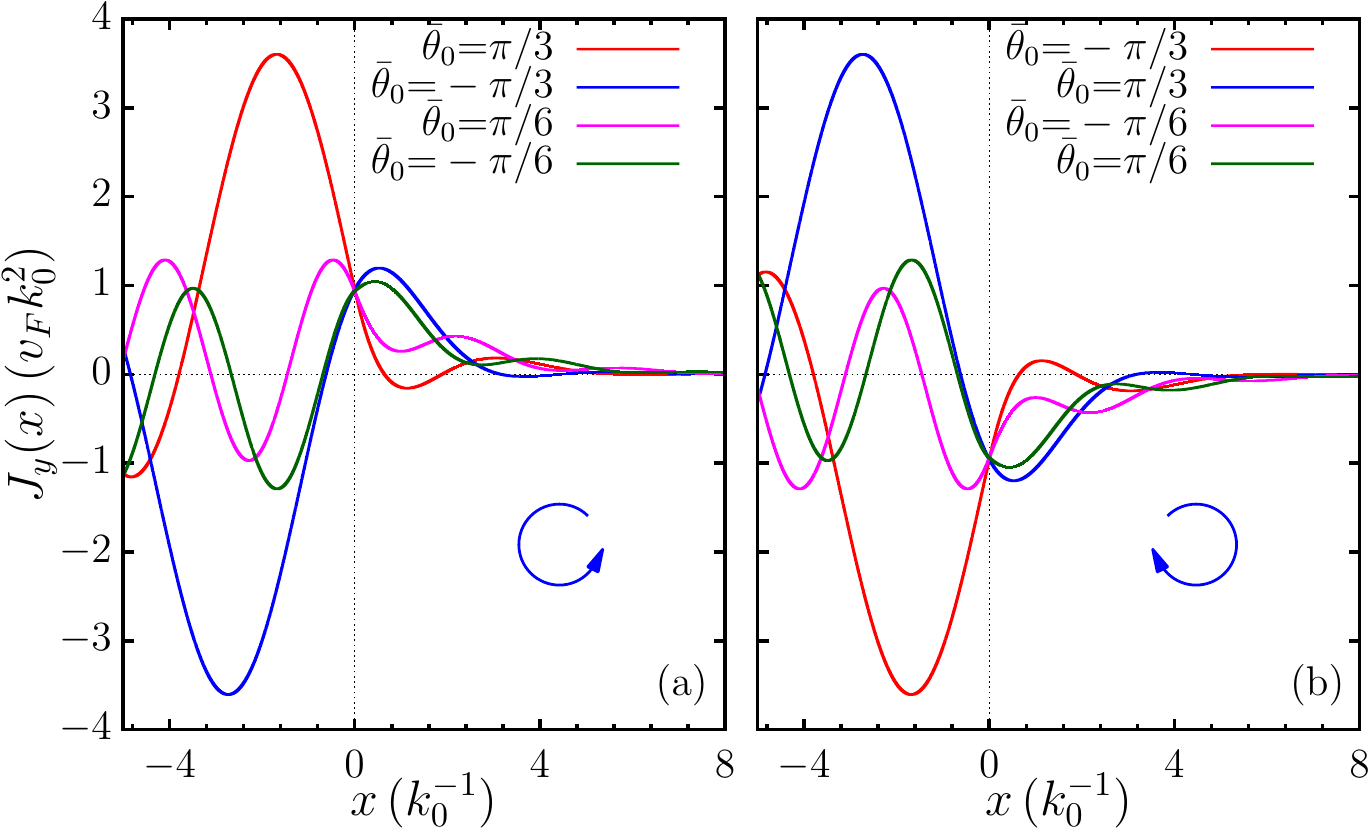}
\caption{$J_y(x)$ component of the current density as a function of the coordinate $x$ (parameter $\eta=0.23$ and $\mu=0$). The line $x=0$ defines the interface with the irradiated region ($x{>}0$). In every case the helicity of the field $\bm A(t)$ is indicated. We see that the direction of the current at the interface depends on the helicity of $\bm A(t)$ and not on the direction of incidence of the electrons \label{currents}.}
\end{center}
\end{figure}

%%%%%%%%%%%%%%%%%%%%%%%%%%%%%%%%%%%%%%%%%%%%%%%%
Besides the reflection and transmission coefficients, it is also interesting to investigate the nature of the currents that appear along the interface between the illuminated and non illuminated regions. As we mentioned before, when $\mu$ lies inside the dynamical gap there is no total transmitted current in the $\hat{\bm{x}}$ direction. Hence, in the irradiated region we can write the current density as  $\bm{J}(x)=J_y(x)\,\hat{\bm{y}}$ ---the translation symmetry along the $y$ axis implies that the current depends only on $x$. Here, $J_y(x)$ can be shown to have the general form
%%%%%%%%%%%%%%%%%%%%%%%%%%%%%%%%%%%%%%%%%%%%%%%%%%%%%%%%%%
\begin{equation}\label{ycurrent}
  J_y(x)=e^{-2\beta x}(A+B_c\,\cos2\alpha x+B_s\,\sin2\alpha x) \,,
\end{equation}
%%%%%%%%%%%%%%%%%%%%%%%%%%%%%%%%%%%%%%%%%%%%%%%%%%%%%%%%%%
$\alpha$, $\beta$ are obtained from $k_+=\alpha{+}i\beta$; $A$, $B_c$ and $B_s$  being constants depending on $\theta_0$, $\eta$ and $\mu$ (and eventually on the laser's polarization). Figure~\ref{currents} shows plots of $J_y(x)$ for different values of the incidence angle ${\theta}_0$ and the two possible circular polarizations of the vector field $\bm A(t)$ for  $\mu=0$ (center of the dynamical gap). Clearly,  $J_y(x)$ is exponentially localized near the  border for $x>0$, its global direction being ultimately determined by the polarization of the vector field  and not by the direction of the incident electron wave. We then refer to these currents as {\it chiral} currents. The comparison between the two polarization makes evident that $J_y^{+}(x, \theta_0)=-J_y^{-}(x,-\theta_0)$, where the $+$ ($-$) sign here refers to the clockwise (counterclockwise) orientation of the polarization. This contrasts with the undriven case where the direction of these currents depends on the incident angle $\theta_0$.

\begin{figure}[t]
\begin{center}
\includegraphics[width=0.9\columnwidth]{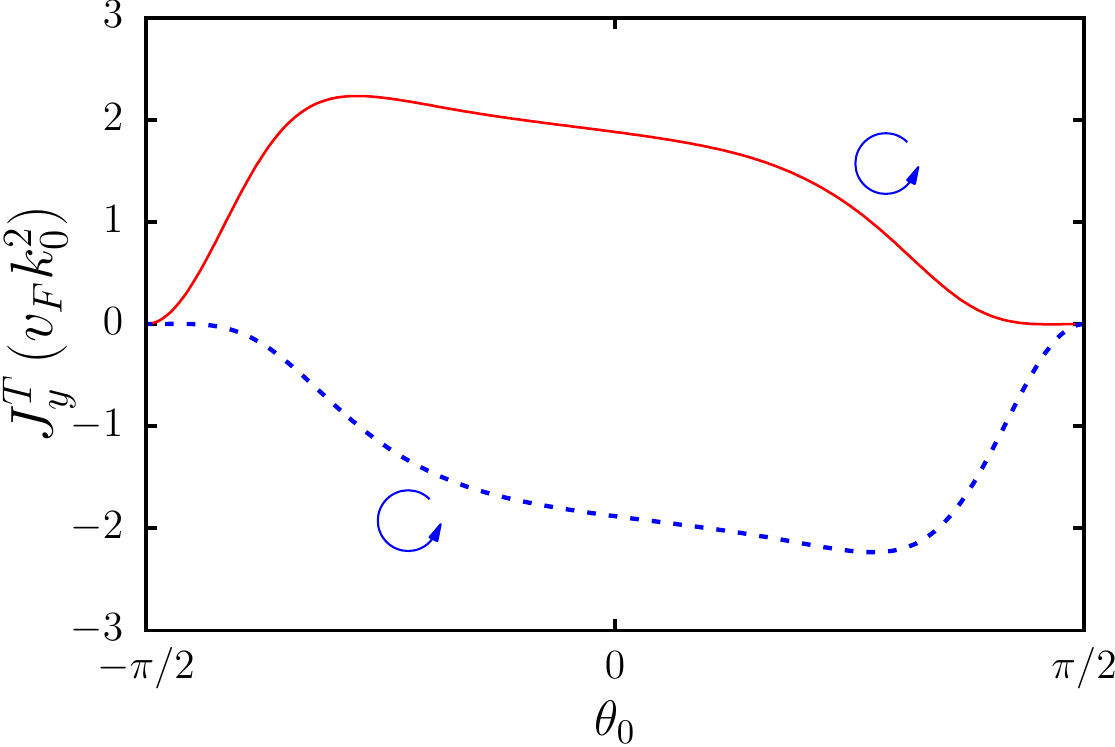}
\caption{ Integrated $J_y$ current in the irradiated region  as a function of the angle of incidence $\theta_0$. It is obvious that for normal incidence ($\theta_0$)  there is a total current along the interface. Here is more visible the relation $J_y^{+}(x, \theta_0)=-J_y^{-}(x,-\theta_0)$ (see text) \label{current2}.}
\end{center}
\end{figure}

It can be shown that even at normal incidence there is a current along the interface equal to $-4|r_0|\sin\varphi_0$, where $r_0=|r_0|\,e^{i\varphi_0}$. This is again a consequence of the chirality imposed by the laser field. This effect can be better appreciated through the integrated current $J_y^T(\theta_0)=\int_0^{+\infty} J_y(x)\,dx$ shown in Fig.~\ref{current2} for $\mu=0$. 
%%%%%%%%%%%%%%%%%%%%%%%%%%%%%%%%%%%%%%%%%%%%%%%%%%%%%%%%%%
%%%%%%%%%%%%%%%%%%%%%%%%%%%%%%%%%%%%%%%%%%%%%%%%%%%%%%

%%%%%%%%%%%%%%%%%%%%%%%%%%%%%%%%%%%%%%%%%%%%%%%%%%%%%%%%%%
%%%%%%%%%%%%%%%%%%%%%%%%%%%%%%%%%%%%%%%%%%%%%%%
\section{Anomalous Goos-H\"anchen Shift}\label{goosH}%%%%%%%
%%%%%%%%%%%%%%%%%%%%%%%%%%%%%%%%%%%%%%%%%%%%%%%
When one considers the scattering of a beam with a finite-size section instead of extended  plane waves interesting effects might arise. One of them involves the reflected  beam being  shifted along the direction of the reflective interface, an effect  known as the Goos-H\"anchen shift. This phenomenon  was originally studied for the case of a light beam in the problem of total internal reflection~\cite{Goos1947} and it was recently considered for the case of electrons in graphene \cite{Beenakker2009} without any driving field. As we show below, owed to the presence of the driving, the Goos-H\"anchen shift becomes chiral.  

A finite size electron beam impinging obliquely onto the interface  can be constructed by an appropriate superposition of plane waves and their corresponding spinors. For the incident beam in the $m=0$ channel we have 
%%%%%%%%%%%%%%%%%%%%%%%%%%%%%%%%%%%%%%%%%%%%%%%%%%%%%%%%%%
\begin{equation}\label{phi0}
\Phi_0^i(x,y) = \int_{-\infty}^{\infty} dk_y\, f(k_y{-}\bar k_y)e^{i(k_y y+k_{x}^{(0)}x)} 
\binom{ e^{-i\frac{\theta_0}{2}}}{e^{i\frac{\theta_0}{2}}}\,.
\end{equation}
%%%%%%%%%%%%%%%%%%%%%%%%%%%%%%%%%%%%%%%%%%%%%%%%%%%%%%%%%%
Here the function $f(k_y{-}\bar k_y)$ is peaked around its mean value $\bar k_y$, which written as ${\bar k_y}=k_0^+\sin{\bar \theta_0}$ gives us the incident angle of the beam $\bar \theta_0$. Notice that both $k_{x}^{(0)}=k_0^+\cos\theta_0$ and $\theta_0$ are functions of $k_y$ as we consider an incident beam with a fixed quasienergy. In the following we consider a beam with a Gaussian profile and take $f(k_y{-}\bar k_y){\propto}\exp(-(k_y{-}\bar k_y)^2/2\sigma^2)$. After integrating this two-component spinor at $x=0$ (the interface) one finds that each component is peaked in coordinate space around different points---in the case of graphene this can be viewed as a consequence of the two-atom basis in the crystal structure. This integration is not analytical but can be approximated by expanding $\theta_0(k_y)$ around $\bar k_y$
%%%%%%%%%%%%%%%%%%%%%%%%%%%%%%%%%%%%%%%%%%%%%%%%%%%%%%%%%%
\begin{equation}\label{approx-theta}
  \theta_0(k_y) \approx \theta_0(\bar k_y) + (k_y-\bar k_y)  \,\left(\frac{d\theta_0}{dk_y} \right)_{\bar k_y}.
\end{equation}
%%%%%%%%%%%%%%%%%%%%%%%%%%%%%%%%%%%%%%%%%%%%%%%%%%%%%%%%%%
With this linear approximation it is straightforward to verify that the two components of $\Phi_0^i(0,y)$ are Gaussian functions centered  at points $y_0^A=-y_0^B=1/2\,(d\theta_0/dk_y)_ {k_y=\bar k_y}$. Therefore, the mean position of the incident beam at the interface, defined as $\bar y_0^i=(y_0^A{+}y_0^B)/2$, is zero.  This result, that means that the incident beam reaches the interface at $y=0$ , relies on the choice of phases of our spinor in Eq.~\eqref{phi0}, but does not alter in any way the final result of the Goos-H\"anchen shift (which will be measured relative to this point).

%%%%%%%%%%%%%%%%%%%%%%%%
%%%%%%%%%%%%%%%%%%%%

A similar treatment can be carried out for the reflected beams in each	 channel. In this case, we have to introduce the reflection coefficients $r_0(k_y)$ and $r_1(k_y)$, and the angle $\theta_1$ for the direction of the reflected beam in  channel $m=1$
%%%%%%%%%%%%%%%%%%%%%%%%%%%%%%%%%%%%%%%%%%%%%%%%%%%%%%%%%%
\bea
\nonumber
\Phi_{1}^r(x,y) &\!=\!&\int_{-\infty}^{\infty} dk_y\, r_1\, f(k_y-\bar k_y) e^{i(k_y y+k_x^{(1)}x)}  \binom{-e^{-i\frac{\theta_1}{2}}}{e^{i\frac{\theta_1}{2}}}\,, \\
\nonumber
\Phi_{0}^r(x,y) &\!=\!&\int_{-\infty}^{\infty} dk_y\, r_0\, f(k_y-\bar k_y)  e^{i(k_y y-k_x^{(0)}x)} \binom{-e^{i\frac{\theta_0}{2}}}{e^{-i\frac{\theta_0}{2}}}\,.\\
\label{pre01}
\eea
%%%%%%%%%%%%%%%%%%%%%%%%%%%%%%%%%%%%%%%%%%%%%%%%%%%%%%%%%%
Writing the reflection coefficients as $r_0=|r_0|\,e^{i\varphi_0}$ and $r_1=|r_1|\,e^{i\varphi_1}$, we can proceed as before and expand the phases up to a linear term 
%%%%%%%%%%%%%%%%%%%%%%%%%%%%%%%%%%%%%%%%%%%%%%%%%%%%%%%%%%
\bea
 \varphi_0(k_y) &\approx& \varphi_0(\bar k_y) + (k_y-\bar k_y) \,\left(\frac{d\varphi_0}{dk_y} \right)_{\bar k_y} \,,\notag\\
 \varphi_1(k_y) &\approx& \varphi_1(\bar k_y) + (k_y-\bar k_y) \,\left(\frac{d\varphi_1}{dk_y} \right)_{\bar k_y}\,.  
\eea
%%%%%%%%%%%%%%%%%%%%%%%%%%%%%%%%%%%%%%%%%%%%%%%%%%%%%%%%%%
Then, using the approximations $|r_0(k_y)|\approx |r_0(\bar k_y)|$ and $|r_1(k_y)|\approx |r_1(\bar k_y)|$ (due to the sharpness of the function $f(k_y-\bar k_y)$ and assuming the modules of $r_0$ and $r_1$ are smooth enough), we get the mean position of $\Phi_{0}^r(0,y)$ and $\Phi_{1}^r(0,y)$. Namely,
%%%%%%%%%%%%%%%%%%%%%%%%%%%%%%%%%%%%%%%%%%%%%%%%%%%%%%%%%%
\bea
  \bar y_{0}^r & = &-\Big(\frac{d\varphi_0}{dk_y} \Big)_{\bar k_y}\,, \nonumber \\
  \bar y_{1}^r & =& -\Big(\frac{d\varphi_1}{dk_y} \Big)_{\bar k_y}\,.
\eea
%%%%%%%%%%%%%%%%%%%%%%%%%%%%%%%%%%%%%%%%%%%%%%%%%%%%%%%%%%\\
The GH shift of the reflected beams are defined as the differences between their mean position and that of the incident beam, $\delta_{0}^r =\bar y_{0}^r{-}\bar y_0^i$ and $\delta_{1}^r =\bar y_{1}^r{-}\bar y_0^i$. Re-writing them in terms of $\theta_0$-derivatives we finally get
%%%%%%%%%%%%%%%%%%%%%%%%%%%%%%%%%%%%%%%%%%%%%%%%%%%%%%%%%%
\bea\label{deltas}
  \delta_{r0} &=&-\left(\frac{d\varphi_0}{d\theta_0}\right)_{\bar k_y}\left( \frac{d\theta_0}{dk_y} \right)_{\bar k_y} \,,  \nonumber\\
  \delta_{r1} &=&-\left(\frac{d\varphi_1}{d\theta_0}\right)_{\bar k_y}\left(\frac{d\theta_0}{dk_y}\right)_{\bar k_y} \,.
\eea
%%%%%%%%%%%%%%%%%%%%%%%%%%%%%%%%%%%%%%%%%%%%%%%%%%%%%%%%%%
Therefore, the GH shifts depend on the derivative of the phases $\varphi_0$ and $\varphi_1$ with respect to $\theta_0$ and a geometrical factor given by $d\theta_0/dk_y=1/(k_0^+\cos\theta_0)$~\cite{Beenakker2009}. 
%%%%%%%%%%%%%%%%%%%%%%%%%%%%%%%%%%%%%%%%%%%%%%%%%%%%%%%%%%
\begin{figure}[t]
\begin{center}
\includegraphics[width=0.95\columnwidth]{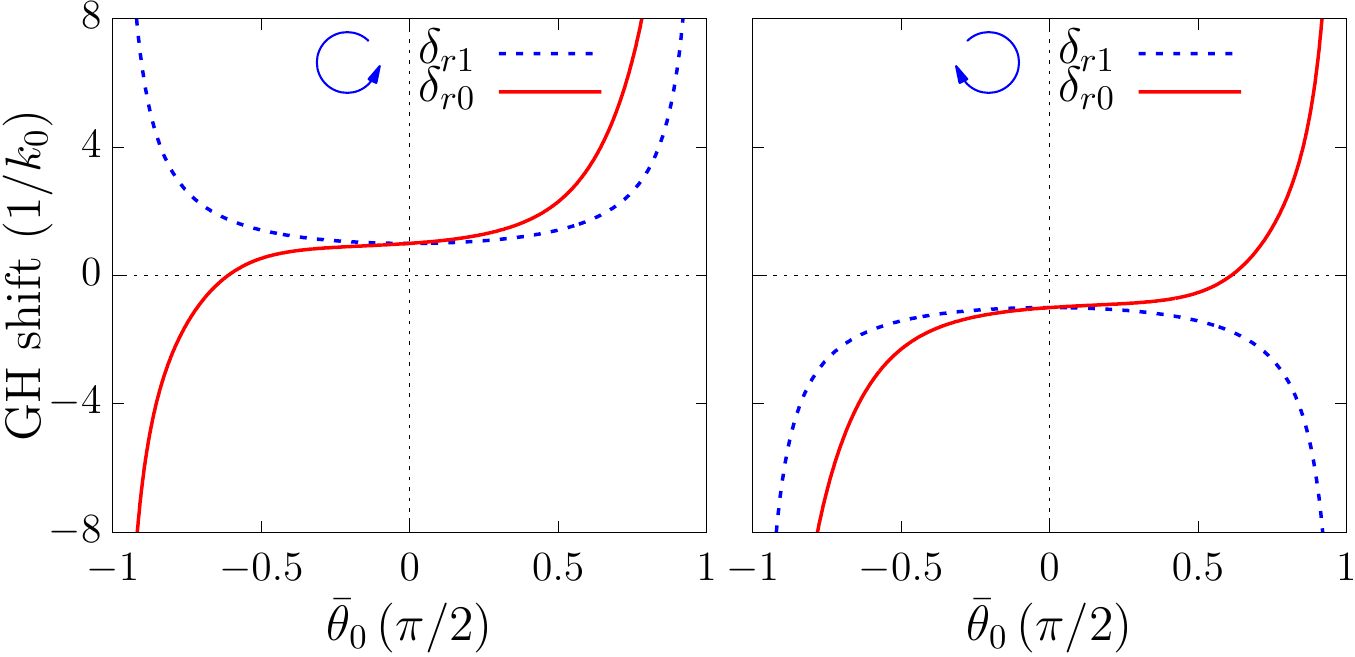}
\caption{Goos-H\"anchen shift of reflected electrons in every channel. The helicity of $\bm A(t)$ is shown on each case. 
The direction of this shift in $m=1$ channel  is independent on the angle of incidence  and depends only on the helicity of the vector field. This is observed only partially in channel $m=0$. There we find a critical angle which marks a change of sign  in this channel, a result similar to that obtained in \cite{Beenakker2009}. The values used were $\eta=0.23$ and $\mu=0$ \label{goos6}.}
\label{shift}
\end{center}
\end{figure}
%%%%%%%%%%%%%%%%%%%%%%%%%%%%%%%%%%%%%%%%%%%%%%%%%%%%%%%%%%
In general, analytical expressions for Eqs.~\eqref{deltas} are rather cumbersome and it is simpler to evaluate them numerically. However, for $\mu=0$ it is straightforward to show that $d \varphi_1/d \theta_0=\pm 1$, where the sign depends only on the helicity of the polarization of the potential vector. This leads to the following expression for the GH shift in the $m=1$ channel
%%%%%%%%%%%%%%%%%%%%%%%%%%%%%%%%%%%%%%%%%%%%%%%%%%%%%%%%%%
\begin{equation}\label{dr1}
  \delta_{r1}(\bar{\theta}_0)=\left\{\begin{array}{l}
    \displaystyle+\frac{1}{k_0\cos\bar{\theta}_0} \qquad\bm{A}(t)\;\mbox{anticlockwise}\,, \\
    \displaystyle-\frac{1}{k_0\cos\bar{\theta}_0} \qquad\bm{A}(t)\;\mbox{clockwise}\,.
  \end{array}
  \right.
\end{equation}
%%%%%%%%%%%%%%%%%%%%%%%%%%%%%%%%%%%%%%%%%%%%%%%%%%%%%%%%%%
(where we have used $k_0^+=k_0$ for $\mu=0$). An important feature is that the GH shift does not depend on $\eta$ (for $\mu=0$) while its sign is independent of $\bar{\theta}_0$. The angle dependence of both $\delta_{r1}(\bar{\theta}_0)$ an $\delta_{r0}(\bar{\theta}_0)$  is shown in Fig. \ref{shift} for both polarizations. We notice that Eqs.~\eqref{dr1} imply that $|\delta_{r1}(\bar{\theta}_0)|=|y_0^A-y_0^B|$ and hence the reflected beam in the $m=1$ channel is shifted exactly as to overlap the maximum of the incident beam in one of the lattices with that of the reflected beam in the opposite one. Namely, either $y_1^{rA}=y_0^{B}$ or $y_1^{rB}=y_0^{A}$.  
It is important to point out that when $\bar{\theta}_0=0$ (normal incidence), there is still a shift given simply by $\pm k_0^{-1}$. This anomalous GH shifts can be related to the appearance of chiral topologically protected interface states as discussed in the next section.

\begin{figure}[t]
\begin{center}
\includegraphics[width=0.9\columnwidth]{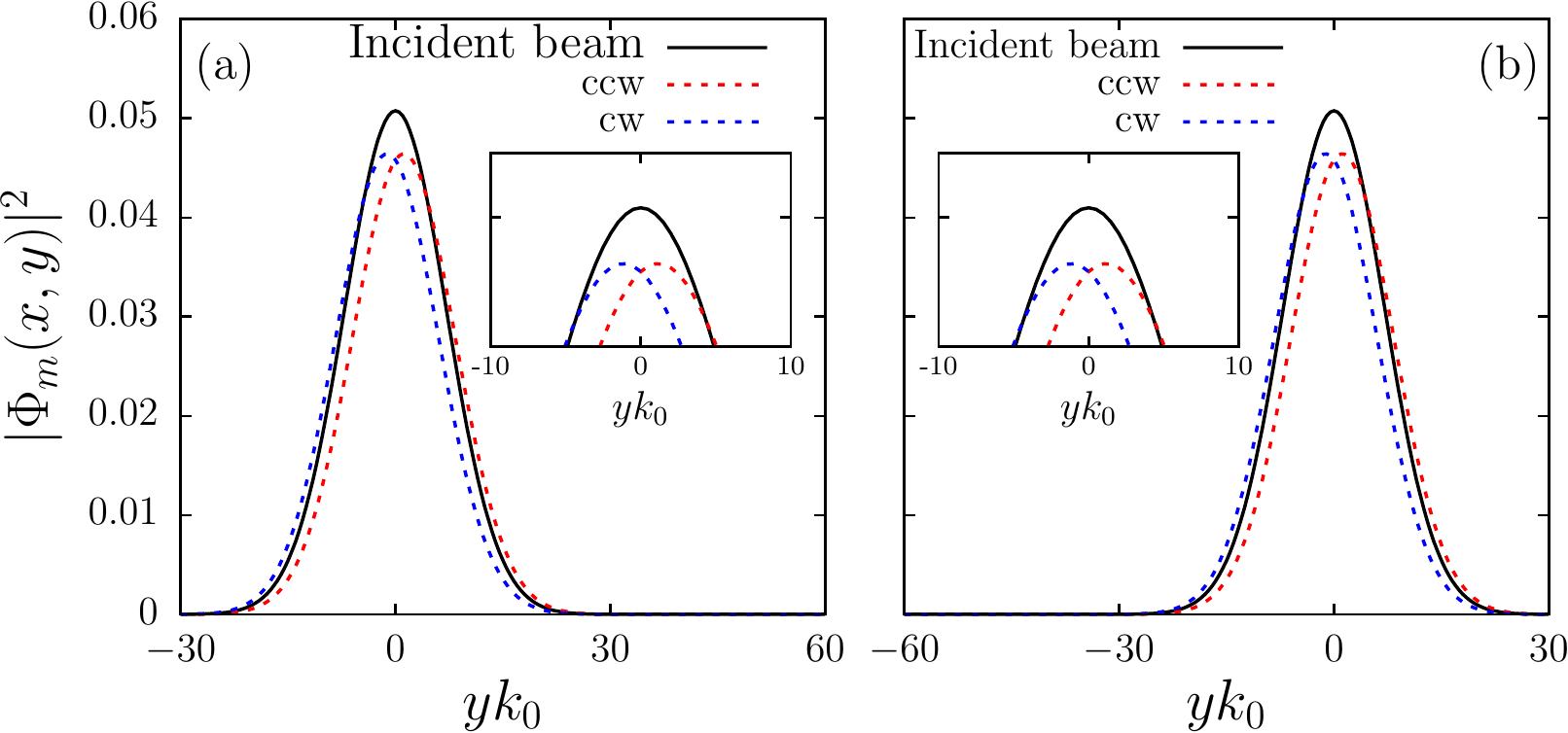}
\caption{(Color online) Profiles as a function of $y$ along the interface ($x=0$) of $|\Phi^i_0(x,y)|^2$ (solid black line) and $|\Phi^r_1(x,y)|^2$ (dashed colored lines), for the two different circular polarizations. (a) $\bar{\theta}_0=+\pi/6$ and (b) $\bar{\theta}_0=-\pi/6$. In either case we see that the counterclockwise (ccw) shift in channel $m=1$ (red line) is larger that the clockwise (cw) one (blue line), as is suggested by Eqs.~\eqref{dr1}, which means that the GH shift here is chiral \label{goos-shift}.}

\end{center}
\end{figure}

To explicitly show the chiral properties of the shift we plot in Fig.~\ref{goos-shift} the profile  along the interface  of the probability density for the incident beam $|\Phi_0^i(x{=}0,y)|^2$ (solid black line) and the scattered beam through the $m=1$ channel $|\Phi_1^r(x{=}0,y)|^2$:  Red (blue) dotted lines for a counterclockwise (clockwise) polarization of the laser field. An incident beam with positive [Fig.~\ref{goos-shift}(a)] and negative [Fig.~\ref{goos-shift}(b)] incident angle is considered. It is clear  that $\delta_{r1}(\bar{\theta}_0)$ has the same sign, independently of $\bar{\theta}_0$, and that $\delta_{r1}^{\mathrm{ccw}}(\bar{\theta}_0)=-\delta_{r1}^{\mathrm{cw}}(\bar{\theta}_0)$,  as stated by Eqs.~\eqref{dr1}.
% [the superscripts stand for counterclockwise ($ccw$) and clockwise ($cw$)].

The behavior of the corresponding shift in the $m=0$ channel is significantly different. Here, while a similar chiral effect exist, there is also a critical value of the incident angle (whose sign is determined also by the helicity of the laser field) where the shift goes to zero and changes sign.

%%%%%%%%%%%%%%%%%%%%%%%%%%%%%%%%%%%%%%%%%%%%%%%%%%%%%%%%%%
%%%%%%%%%%%%%%%%%%%%%%%%%%%%%%%%%%%%%%%%%%%%%%%%%%%%%%%%%%

%%%%%%%%%%%%%%%%%%%%%%%%%%%%%%%%%%%%%%%%%%%%%%%%%%%%%%%%%%
\subsection{Conexion between the GH shift and the presence of topological egde states}
\begin{figure}[t]
\includegraphics[width=0.85\columnwidth]{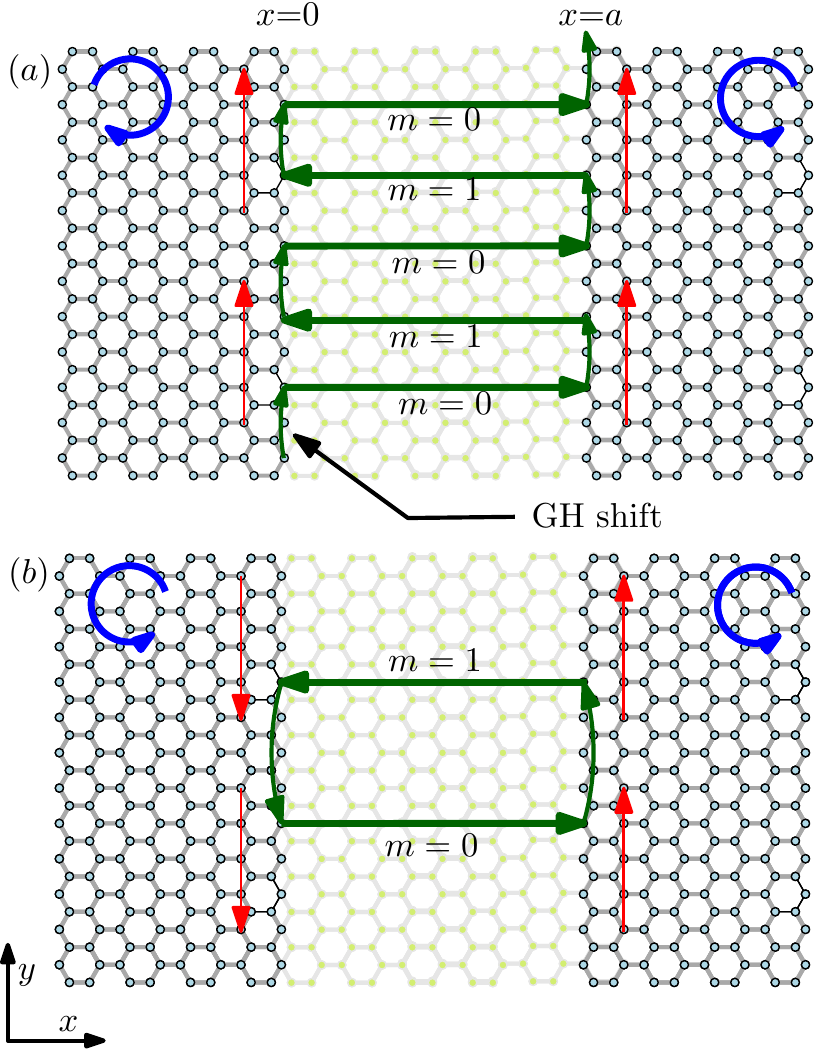}
\caption{(Color online) Schematic representation showing the connection between the GH shif and the topological states. (a) Emergence of chiral states at the interface between two irradiated region with opposite polarizations as the non irradiated zone that separates them shrinks. (b) In the case of equal polarizations the GH shift at each interface cancels out and there are no chiral states. \label{ES}}
\end{figure}

%%%%%%%%%%%%%%%%%%%%%%%%%%%%%%%%%%%%%%%%%%%%%%%%%%%%%%%%%%
It is interesting to look for a connection between the anomalous GH shift and the presence of chiral states due to the nontrivial topological character that the time dependent field introduces.
In the presence of a single interface, like the one we have discussed so far, we have show that an incident plane wave lead to the appearance of chiral current at the interface. In the case of a beam this leads to the GH shift. Let us now consider the situation depicted in Fig.~\eqref{ES}(a). This corresponds to two irradiated regions separated by a non irradiated one of width $a$ (not to be confused with the lattice constant). The circular polarization of the laser fields on each region is the opposite. Hence, as the two regions have opposite topological numbers, chiral edge states must exist between them  as the non irradiation region shrinks to zero~\cite{Calvo2015}. To see how this is related to the anomalous GH shift, we consider an incident beam in the $m=0$ channel that gets reflected and shifted in the $m=1$ channel  at the right interface (see Fig.~\eqref{ES}(a)). This reflected beam in turn is reflected back at the left interface in the $m=0$ channel but GH shifted in the \textit{same} direction as the polarization of the laser field in that interface is the opposite. The cycle repeats and one finds with this simple argument that there must be states in the non irradiated area with a net velocity along the interface. This is in fact what one obtains by solving the full set of equations: There are a number of states inside the non irradiated area with a net chirality (Chern number) of $2$ as expected~\cite{Calvo2015}. The number of such states depends on $a$ and only $2$ survive in the limit $a\rightarrow0$. One can estimate the group velocity $v_g$ of such states as $\delta_1(0)/\tau$ where $\tau=2\hbar/\Delta$ is the time the beam spends in the irradiated region and $\Delta=\eta\hbar\Omega$ is the size of the dynamical gap. This gives
%%%%%%%%%%%%%%%%%%%%%%%%%%%%%%%
\begin{equation}
v_g=\frac{\delta_{r1}(0)}{\tau}=\frac{\Delta}{2\hbar k_0}=2\eta v_F
\end{equation}  
%%%%%%%%%%%%%%%%%%%%%%%%%%%
which is the expected velocity of a edge state crossing the dynamical gap.

The situation is completely different when the  polarization in the two regions is the same (Fig.~\eqref{ES}(b)) as in that case the GH shift on each interface cancel each other leading to a zero group velocity. In particular, one can show that there are subgap states that do not cross the gap and that disappear as $a\rightarrow0$.

%%%%%%%%%%%%%%%%%%%%%%%%%%%%%%%%%%%%%%%%%%%%%%%%%%%%%%%%%%
%%%%%%%%%%%%%%%%%%%%%%%%%%%%%%%%%%%%%%%%%%%%%%%%%%%%%%%%%%

%%%%%%%%%%%%%%%%%%%%%%%%%%%%%%%%%%%%%%%%%%%%%%%%%%%%%%%%%%
\section{Scattering by Circular Obstacles }
%%%%%%%%%%%%%%%%%%%%%%%%%%%%%%%%%%%%%%%%%%%%%%%%%%%%%%%%%%
We now consider the problem of Floquet scattering by a circular irradiated region  of radius $R$ (large in comparison with the lattice constant $a$). As we are again interested in quasienergies in the range of the dynamical gap, we will use the same approximation as in the previous section and retain  only the replicas in the  $m=0$ and $m=1$ channels.
For a better treatment of the problem we will split it into two parts. First, we will solve the scattering problem for an incident wave with circular symmetry and later treat the case of a  plane wave and a beam. 
%%%%%%%%%%%%%%%%%%%%%%%%%%%%%%%%%%%%%%%%%%%%%%%%%%%%%%%%%%
\subsection{Solutions for circularly symmetric incident waves}\label{Circularly Symmetric Incident Waves}
%%%%%%%%%%%%%%%%%%%%%%%%%%%%%%%%%%%%%%%%%%%%%%%%%%%%%%%%%%
To find  the solution in this case, it is better to switch to polar coordinates and redefine the $p_{\pm}$ differential operators that enter the Hamiltonian in terms of those coordinates. Hence, we have
%%%%%%%%%%%%%%%%%%%%%%%%%%%%%%%%%%%%%%%%%%%%%%%%%%%%%%%%%%
\begin{eqnarray}
\label{property}
\nonumber
\pmas &=& \hbar e^{i\theta} (-i\partial_r+\frac{1}{r}\partial_\theta)\,,\\
\nonumber
\pmenos &=& \hbar e^{-i\theta}(-i\partial_r-\frac{1}{r}\partial_\theta)\,,          \\
\pmenos\pmas &=&-\hbar^2\big[\partial_r^2+\frac{1}{r}\partial_r+\frac{1}{r^2}\partial_\theta^2\big]\,.
\end{eqnarray}
%%%%%%%%%%%%%%%%%%%%%%%%%%%%%%%%%%%%%%%%%%%%%%%%%%%%%%%%%%
where the latter equality is defined for later convenience.

%%%%%%%%%%%%%%%%%%%%%%%%%%%%%%%%%%%%%%%%%%%%%%%%%%%%%%%%%%
\begin{figure}[t]
\includegraphics[width=0.7\columnwidth]{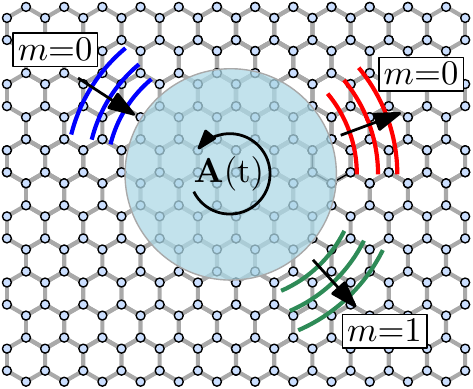}
\caption{Schema of the circularly symmetric scattering: Incident particles in the  $m=0$ channel (blue lines) are scattered  in both the  $m=1$ (red lines) and  $m=0$ (green lines) channels. \label{setup_cyl}}
\end{figure}
%%%%%%%%%%%%%%%%%%%%%%%%%%%%%%%%%%%%%%%%%%%%%%%%%%%%%%%%%%

%%%%%%%%%%%%%%%%%%%%%%%%%%%%%%%%%%%%%%%%%%%%%%%%%%%%%%%%%%
\subsubsection*{Outside the irradiated region}
%%%%%%%%%%%%%%%%%%%%%%%%%%%%%%%%%%%%%%%%%%%%%%%%%%%%%%%%%%
Outside the irradiated circular spot the Floquet replicas are decoupled and so we have to solve the two independent  matrix equations
%%%%%%%%%%%%%%%%%%%%%%%%%%%%%%%%%%%%%%%%%%%%%%%%%%%%%%%%%%
\begin{eqnarray}
\nonumber
  \left(
  \begin{array}{cc}
      \hbar\Omega & \vf\pmenos \\
      \vf \pmas     & \hbar\Omega
  \end{array}    
   \right)
     \binom{u_A}{u_B}
  &=&\varepsilon 
   \binom{u_A}{u_B}\,,   \\
     \left(
  \begin{array}{cc}
      0 & \vf \pmenos \\
      \vf \pmas     & 0
  \end{array}    
   \right)
 \binom{w_A}{w_B}&=&\varepsilon 
\binom{w_A}{w_B}\,.
\end{eqnarray}
%%%%%%%%%%%%%%%%%%%%%%%%%%%%%%%%%%%%%%%%%%%%%%%%%%%%%%%%%%
Notice that only two functions need to be determined as, for instance, the solutions $u_A$ and $w_B$ can be written directly in terms of $u_B$ and $w_A$
%%%%%%%%%%%%%%%%%%%%%%%%%%%%%%%%%%%%%%%%%%%%%%%%%%%%%%%%%%
\begin{eqnarray}
\nonumber
\label{16}
u_A &=& -\frac{\vf \pmenos u_B}{\hbar\Omega-\varepsilon}\,, \\
w_B &=& \frac{\vf \pmas w_A}{\varepsilon}\,.
\end{eqnarray}
%%%%%%%%%%%%%%%%%%%%%%%%%%%%%%%%%%%%%55
Taking advantage of the circular symmetry of the problem we can write $u_B(r,\theta)=e^{in\theta}f(k_0r)$ and  $w_A(r,\theta)=e^{il\theta}g(k_0r)$ where $n$ and $l$ are integer numbers---they are in principle independent but they will be forced to be  equal once the interior of the irradiated region is considered, so we take $n=l$ hereon. It is straightforward to check that $f(x)$ and $g(x)$ satisfy  
%%%%%%%%%%%%%%%%%%%%%%%%%%%%%%%%%%%%%%%%%%%%%%%%%%%%%%%%%
\begin{eqnarray}
\nonumber
\frac{d^2f(x)}{dx^2}+\frac{1}{x}\frac{df(x)}{dx}+\left[(1{-}\mu)^2-\frac{l^2}{x^2}\right]f(x)&=&0 \,, \\
\frac{d^2g(x)}{dx^2}+\frac{1}{x}\frac{dg(x)}{dx}+\left[(1{+}\mu)^2-\frac{l^2}{x^2}\right]g(x)&=&0\, .
\end{eqnarray}
%%%%%%%%%%%%%%%%%%%%%%%%%%%%%%%%%%%%%%%%%%%%%%%%%%%%%%%%%%
where we defined a dimensionless radial coordinate $x=k_0r$.
These are Bessel equations and hence the  solutions can be written as a superposition of Hankel functions of the first ($H_l^{(1)}$) and second ($H_l^{(2)}$) kind
%%%%%%%%%%%%%%%%%%%%%%%%%%%%%%%%%%%%%%%%%%%%%%%%%%%%%%%%%%
\bea
\nonumber
u_B(r>R,\theta) &=&e^{il\theta}\left[A_l H^{(1)}_l(k_0^-r) + B_l H^{(2)}_l(k_0^-r)\right]\,, \\
\nonumber
w_A(r>R,\theta) &=&e^{il\theta}\left[M_l  H^{(1)}_l(k_0^+r) + N_l H^{(2)}_l(k_0^+r)\right]\,,\\
\eea
%%%%%%%%%%%%%%%%%%%%%%%%%%%%%%%%%%%%%%%%%%%%%%%%%%%%%%%%%%
with $k_0^{\pm}=k_0(1\pm\mu)$.
The remaining solutions $u_A$ and $w_B$ are obtained using Eqs.~\eqref{16}.
%%%%%%%%%%%%%%%%%%%%%%%%%%%%%%%%%%%%%%%%%%%%%%%%%%%%%%%%%%
Written all of them as a spinor on each channel, we finally get
%%%%%%%%%%%%%%%%%%%%%%%%%%%%%%%%%%%%%%%%%%%%%%%%%%%%%%%%%%
\bea\label{fullsolutions}
\nonumber
\Phi_1^{(l)}(r>R,\theta) &= &A_l\,e^{il\theta} \left( \begin{array}{c}
i e^{-i\theta} H_{l-1}^{(1)}(k_0^-r) \\
H_{l}^{(1)}(k_0^-r)
\end{array} \right) \\
\nonumber
&+&
B_l\,e^{il\theta} \left( \begin{array}{c}
i e^{-i\theta} H_{l-1}^{(2)}(k_0^-r) \\
H_{l}^{(2)}(k_0^-r)
\end{array} \right) \,, \\ %%%%%%%%%%%%%%%
\nonumber
\Phi_0 ^{(l)}(r>R,\theta) &=& M_l\,e^{il\theta} \left( \begin{array}{c}
H_l^{(1)}(k_0^+r) \\
i e^{i\theta}H_{l+1}^{(1)}(k_0^+r)
\end{array} \right) \\
&+&
N_l\,e^{il\theta} \left( \begin{array}{c}
H_l^{(2)}(k_0^+r) \\
i e^{i\theta}H_{l+1}^{(2)}(k_0^+r) \\
\end{array} \right) \,.
\eea
%%%%%%%%%%%%%%%%%%%%%%%%%%%%%%%%%%%%%%%%%%%%%%%%%%%%%%%%%%
We choose the incoming electrons to be only in the  $m=0$ channel, which implies that $A_l=0$. This can be verified by looking at the radial component of the probability current $J_r=\Phi^\dagger\bm{\sigma}\cdot \hat{\bm{r}}\Phi=2\,\mathrm{Re}[e^{i\theta}\phi_A\phi_B^*]$, with $\hat{\bm{r}}=(\cos\theta,\sin\theta)$ and $\Phi=(\phi_A,\phi_B)^T$. Taking into account that for $x\gg1$ 
%%%%%%%%%%%%%%%%%%%%%%%%%%%%%%%%%%%%%%%%%%%%%%%%%%%%%%%%%% 
\bea
\label{asinth}
\nonumber
  H_l^{(1)}(x) &\simeq& \sqrt{\frac{2}{\pi x}}\,e^{i(x-(l+\frac{1}{2})\frac{\pi}{2})}\,, \\
  H_l^{(2)}(x) &\simeq& \sqrt{\frac{2}{\pi x}}\,e^{-i(x-(l+\frac{1}{2})\frac{\pi}{2})}\, ,
\eea
%%%%%%%%%%%%%%%%%%%%%%%%%%%%%%%%%%%%%%%%%%%%%%%%%%%%%%%%%%
it is easy to verify that the first (second) spinor in  $\Phi^{(l)}_1$  is an incoming (outgoing) wave. Similarly,  the first (second) spinor in $\Phi^{(l)}_0$ is an outgoing (incoming) wave. 

%%%%%%%%%%%%%%%%%%%%%%%%%%%%%%%%%%%%%%%%%%%%%%%%%%%%%%%%%%
\subsubsection*{Inside the irradiated region}
%%%%%%%%%%%%%%%%%%%%%%%%%%%%%%%%%%%%%%%%%%%%%%%%%%%%%%%%%%
The procedure in this case is basically the same as before, except that now the two Floquet channels are coupled. 
Writing  $u_{B}(r,\theta)=e^{il\theta}F(k_0r)$ and $w_{A}(r,\theta)=e^{il\theta}G(k_0r)$, we get the following equations
%%%%%%%%%%%%%%%%%%%%%%%%%%%%%%%%%%%%%%%%%%%%%%%%%%%%%%%%%%
\bea
\nonumber
\frac{d^2F}{dx^2}{+}\frac{1}{x}\frac{dF}{dx}+\left[(1{-}\mu)^2{-}\frac{l^2}{x^2}\right]F &=&-2\eta(1{-}\mu)G, \\
\frac{d^2G}{dx^2}{+}\frac{1}{x}\frac{dG}{dx}+\left[(1{+}\mu)^2{-}\frac{l^2}{x^2}\right]G &=&2\eta(1{+}\mu)F.
\eea
%%%%%%%%%%%%%%%%%%%%%%%%%%%%%%%%%%%%%%%%%%%%%%%%%%%%%%%%%%
This system of equations can be solved  by using the substitutions $F(x)=C\,I_l(\lambda x)$ and $G(x)=D\,I_l(\lambda x)$, $C$ and $D$ being integration constants  and $I_l$ the  modified Bessel functions of the second kind (which are well behaved at $r=0$). This leads to a secular equation for $\lambda$ 
%%%%%%%%%%%%%%%%%%%%%%%%%%%%%%%%%%%%%%%%%%%%%%%%%%%%%%%%%%
\begin{equation}
\left[\lambda^2+(1-\mu)^2\right]\left[\lambda^2+(1+\mu)^2\right]+4\eta^2\left(1-\mu^2\right)=0.
\end{equation}
%%%%%%%%%%%%%%%%%%%%%%%%%%%%%%%%%%%%%%%%%%%%%%%%%%%%%%%%%%
Inside the dynamical gap, the possible values  for $\lambda$ have the form $\pm\left(a\pm ib\right)$. From these four values only two give different solutions. Hence, it will suffice to take only the two conjugate solutions $\lambda_-=\lambda_+^*$ with positive real part
%%%%%%%%%%%%%%%%%%%%%%%%%%%%%%%%%%%%%%%%%%%%%%%%%%%%%%%%%%
\begin{equation}
\lambda_{+}=\sqrt{-1-\mu^2 + 2i\sqrt{\eta^2-\mu^2(1+\eta^2)}},
\end{equation}
%%%%%%%%%%%%%%%%%%%%%%%%%%%%%%%%%%%%%%%%%%%%%%%%%%%%%%%%%%
the square root for $\lambda_+$ is taken in the principal branch. The solutions for $u_{B}$ and $w_{A}$ are then
%%%%%%%%%%%%%%%%%%%%%%%%%%%%%%%%%%%%%%%%%%%%%%%%%%%%%%%%%%
\bea
\nonumber
u_{B} (r\leq R,\theta)&=&e^{il\theta} \left[ d_{l}^+ \Pi_+ I_l(\kappa_0^+r)\!+\!d_{l}^- \Pi_- I_l(\kappa_0^-r)\right]\,, \\
 w_{A}(r\leq R,\theta) &=&e^{il\theta} \left[ d_{l}^+ I_l(\kappa_0^+r)+d_{l}^- I_l(\kappa_0^-r)\right],
\eea
%%%%%%%%%%%%%%%%%%%%%%%%%%%%%%%%%%%%%%%%%%%%%%%%%%%%%%%%%%
with $\kappa_0^{\pm}=\lambda_\pm k_0$, $\Pi_{\pm}=(\lambda_{\pm}^2{+}(1{+}\mu)^2)/2\eta(1{+}\mu)$ and $d_{l}^\pm$ being integration constants. The final solutions in a spinor form can be written as follows
%%%%%%%%%%%%%%%%%%%%%%%%%%%%%%%%%%%%%%%%%%%%%%%%%%%%%%%%%%
\bea\label{interior_plane}
\nonumber
\Phi_{1}^{(l)}(r\leq R,\theta)&=&e^{il\theta}d_{l}^+ \Pi_+\left( \begin{array}{c}
\frac{ie^{-i\theta}}{1-\mu}\lambda_+I_{l-1}(\kappa_0^+r) \\
I_l(\kappa_0^+r)
\end{array} \right) \\
\nonumber
&+&  e^{il\theta}d_{l}^- \Pi_-\left( \begin{array}{c}
\frac{ie^{-i\theta}}{1-\mu}\lambda_-I_{l-1}(\kappa_0^-r) \\
I_l(\kappa_0^-r)
\end{array} \right), \\ %%%%%%%%%%%%%%%
\nonumber
\Phi_{0}^{(l)}(r\leq R,\theta)&=&e^{il\theta}d_{l}^+ \left( \begin{array}{c}
I_{l}(\kappa_0^+r) \\
-\frac{ie^{i\theta}}{1+\mu}\lambda_+I_{l+1}(\kappa_0^+r)
\end{array} \right) \\
&+&  e^{il\theta}d_{l}^- \left( \begin{array}{c}
I_{l}(\kappa_0^-r) \\
-\frac{ie^{i\theta}}{1+\mu}\lambda_-I_{l+1}(\kappa_0^-r)
\end{array} \right).
\eea
%%%%%%%%%%%%%%%%%%%%%%%%%%%%%%%%%%%%%%%%%%%%%%%%%%%%%%%%%%
The complete solution of the problem requires to match the wavefunctions at $r=R$ and so determine the relation between the different integration constant. We will not pursue that here since we will directly use these results as an intermediate  step to solve the more interesting problem of incident plane wave in the next section.
%%%%%%%%%%%%%%%%%%%%%%%%%%%%%%%%%%%%%%%%%%%%%%%%%%%%%%%%%%
\begin{figure}[t]
\begin{center}
\includegraphics[width=0.9\columnwidth]{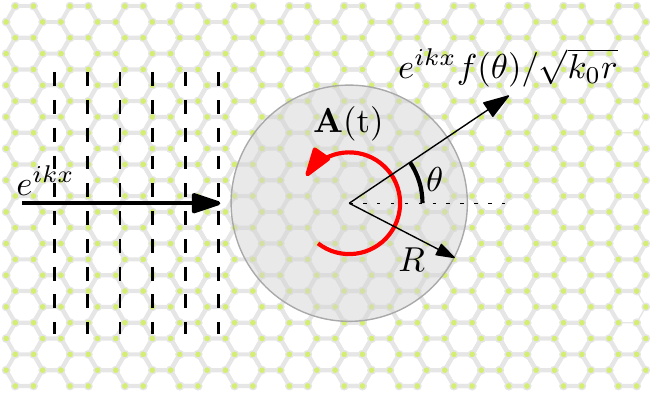}
\caption{Scattering of plane waves. The incident wave is entirely in the $m=0$ channel. The scattered waves have component in each channel $m=0$ and $m=1$. }
\end{center}
\end{figure}
%%%%%%%%%%%%%%%%%%%%%%%%%%%%%%%%%%%%%%%%%%%%%%%%%%%%%
%%%%%%%%%%%%%%%%%%%%%%%%%%%%%%%%%%%%%%%%%%%%%%%%%%%%%%%%%%
\subsection{Incident plane waves}\label{Incident Plane Waves}
%%%%%%%%%%%%%%%%%%%%%%%%%%%%%%%%%%%%%%%%%%%%%%%%%%%%%%%%%%
We now analyze the scattering of a plane wave.  For that, we will consider an homogeneous flux of electrons in the $m=0$ channel represented by the plane wave $\Phi_{\mathrm{inc}}(\bm {r})=e^{ikx}(1,1)^\mathrm{T}/\sqrt{2}$. To take advantage of the results presented in the previous section, we will solve this problem in polar coordinates. To this end, it is useful to expand $\Phi_{\mathrm{inc}}$ in a series of Bessel functions by means of the Jacobi identity
%%%%%%%%%%%%%%%%%%%%%%%%%%%%%%%%%%%%%%%%%%%%%%%%%%%%%%%%%%
\begin{equation}\label{jacobi}
  \Phi_{\mathrm{inc}}(\bm{ r})=\frac{1}{\sqrt{2}}\binom{1}{1}  \sum_{l=-\infty}^{\infty}i^l J_l(k_0^+r)\,e^{il\theta}\,,
\end{equation}
%%%%%%%%%%%%%%%%%%%%%%%%%%%%%%%%%%%%%%%%%%%%%%%%%%%%%%%
where $J_n(x)$ are the Bessel functions of the first kind and we have used the property $k=k_0^+=k_0(1{+}\mu)$. In the outside region (non irradiated), the wavefunction  in the $m=0$ and $m=1$ channels  are obtained in terms of Hankel functions by combining the solutions given by Eqs.~\eqref{fullsolutions} for different $l$ values. In the $m=0$  channel an outgoing scattered solution must be added to the incident one. On the contrary, in the  $m=1$ channel there is only an outgoing wave. Namely
%%%%%%%%%%%%%%%%%%%%%%%%%%%%%%
\bea\label{ext0}
\Phi_1 (r\!>\!R,\theta) & = &\sum_{l=-\infty}^{\infty} r_1^{(l)}e^{il\theta}
  \left(\begin{array}{c}
    i\,e^{-i\theta} H_{l-1}^{(2)}(k_0^-r) \\
    H_l^{(2)}(k_0^-r)
    \end{array} \right)\,, \\
    \nonumber
\Phi_0(r\!>\!R,\theta) & =& \Phi_\mathrm{inc}(\bm{ r})\! +\!\sum_{l=-\infty}^{\infty} r_0^{(l)}e^{il\theta} \left(
\begin{array}{c}
   H_{l}^{(1)}(k_0^+r) \\ ie^{i\theta} H_{l+1}^{(1)}(k_0^+r)
\end{array}  \right)\,.
\eea
%%%%%%%%%%%%%%%%%%%%%%%%%%%%%%
On the other hand, inside the irradiated region the wavefunction is written as a linear combination of the solutions given by Eqs.~\eqref{interior_plane}.
After matching the inside and outside wavefunctions along the circle $r=R$ all the coefficients can be obtained. In particular, the coefficients $r_0^{(l)}$ and $r_1^{(l)}$ are the ones that determine the angular distribution of the scattered probability flux on each channel. 

%%%%%%%%%%%%%%%%%%%%%%%%%%%%%%%%%%%%%%%%%%%%%%%%%%%%%%%%%
\begin{figure}[t]
\includegraphics[width=0.9\columnwidth]{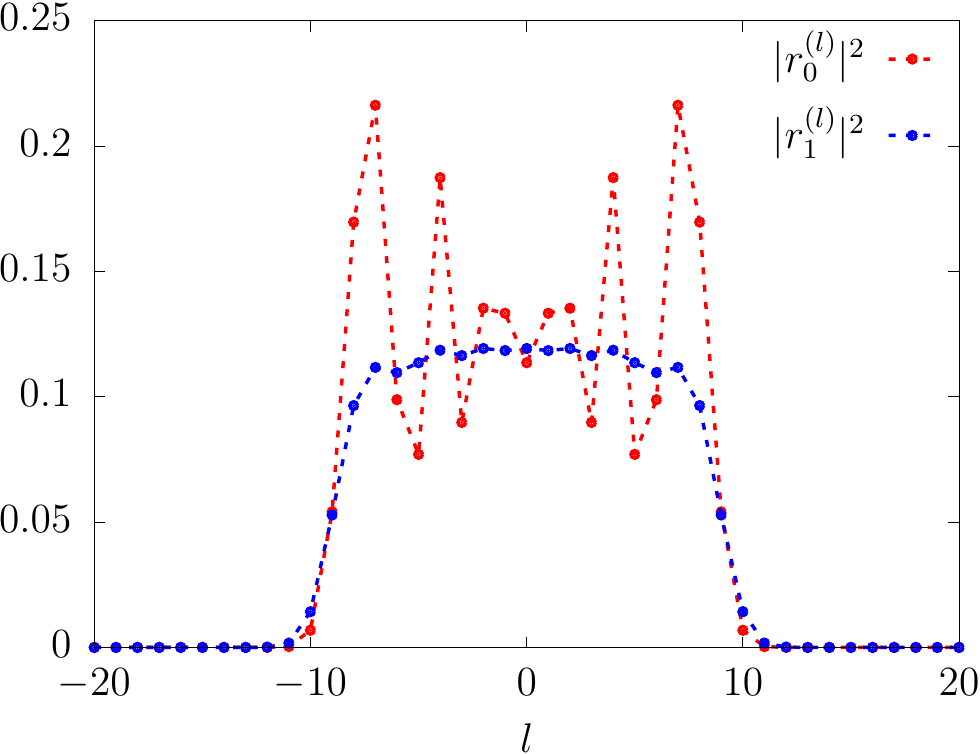}
\caption{Squared module of coefficients  $|r_0^{(l)}|^2$ and  $|r_1^{(l)}|^2$ in the scattering of plane waves as a function of the integer parameter $l$  (see Eqs.~\eqref{ext0}). We see that both coefficients tend to zero when $|l|\gg k_0R$. The calculation was made using $\mu=0$, $\eta=\mbox{0.23}$ and $k_0R=10$.}\label{coeff}
\end{figure}
%%%%%%%%%%%%%%%%%%%%%%%%%%%%%%%%%%%%%%%%%%%%%%%%%%%%%%%%%%%

Fig.~\ref{coeff} shows the obtained values for $|r_0^{(l)}|^2$ and $|r_1^{(l)}|^2$ as a function of $l$  for $\mu=0$ and $k_0R=10$. Clearly, they both vanish when $|l|{\gg} k_0R$. As usual, this can be interpreted using a semiclassical picture where $|l|/k_0$ is a measure of the impact parameter $b$ and hence no scattering is expected for $b>R$. 
%%%%%%%%%%%%%%%%%%%%%%%%%%%%%%%%%%%%%%%%%%%%%%%%%%%%%%%%%%
\subsubsection*{Far field scattering}
%%%%%%%%%%%%%%%%%%%%%%%%%%%%%%%%%%%%%%%%%%%%%%%%%%%%%%%%%%
In order to calculate the angular scattering distribution far from the scattering center we need to describe the wavefuncion's behavior for large $r$. This can be achieved by means of the  asymptotic relations shown in Eqs.~\eqref{asinth}. With these approximations the scattered wavefunctions $\Phi_0^\mathrm{sc}=\Phi_0-\Phi_\mathrm{inc}$  and $\Phi_1$ are
%%%%%%%%%%%%%%%%%%%%%%%%%%%%%%%%%%%%%%%%%%%%%%%%%%%%%%%%%%%%%
\bea
\nonumber	
  \Phi_1 &=&\frac{e^{ik_0^-r}}{\sqrt{r}} f_1(\theta) \frac{1}{\sqrt{2}} \left(
    \begin{array}{c}
       e^{-i\theta/2} \\
       e^{i\theta/2}
    \end{array}
 \right),\\
  \Phi_0^\mathrm{sc} &=&\frac{e^{ik_0^+r}}{\sqrt{r}} f_0(\theta) \frac{1}{\sqrt{2}} \left(
    \begin{array}{c}
       e^{-i\theta/2} \\
       e^{i\theta/2}
    \end{array}
 \right)\,.
\eea
%%%%%%%%%%%%%%%%%%%%%%%%%%%%%%%%%%%%%%%%%%%%%%%%%%%%%%%%%%
The quantities $f_0(\theta)$ and $f_1(\theta)$ are  functions of the coefficients $r_0^{(l)}$ and $r_1^{(l)}$ as follows
%%%%%%%%%%%%%%%%%%%%%%%%%%%%%%%%%%%%%%%%%%%%%%%%%%%%%%%%%%
\bea
  f_1(\theta) &=&\frac{2 e^{-i\theta/2}e^{i\pi/4}}{\sqrt{\pi k_0^-}}  \sum_{l=-\infty}^{\infty} r_1^{(l)}\, e^{il(\theta+\pi/2)}\,, \nonumber\\
  f_0(\theta) &=&\frac{2 e^{i\theta/2}e^{-i\pi/4}}{\sqrt{\pi k_0^+}}  \sum_{l=-\infty}^{\infty} r_0^{(l)}\, e^{il(\theta-\pi/2)}\,.
\eea
%%%%%%%%%%%%%%%%%%%%%%%%%%%%%%%%%%%%%%%%%%%%%%%%%%%%%%%%%%
From these, we can obtain the differential cross sections for every channel, which are defined in the usual way as $d\sigma_m/d\theta=|f_m(\theta)|^2$.
%%%%%%%%%%%%%%%%%%%%%%%%%%%%%%%%%%%%%%%%%%%%%%%%%%%%%%%%%%
\begin{figure}[t]
\begin{center}
\includegraphics[width=0.9\columnwidth]{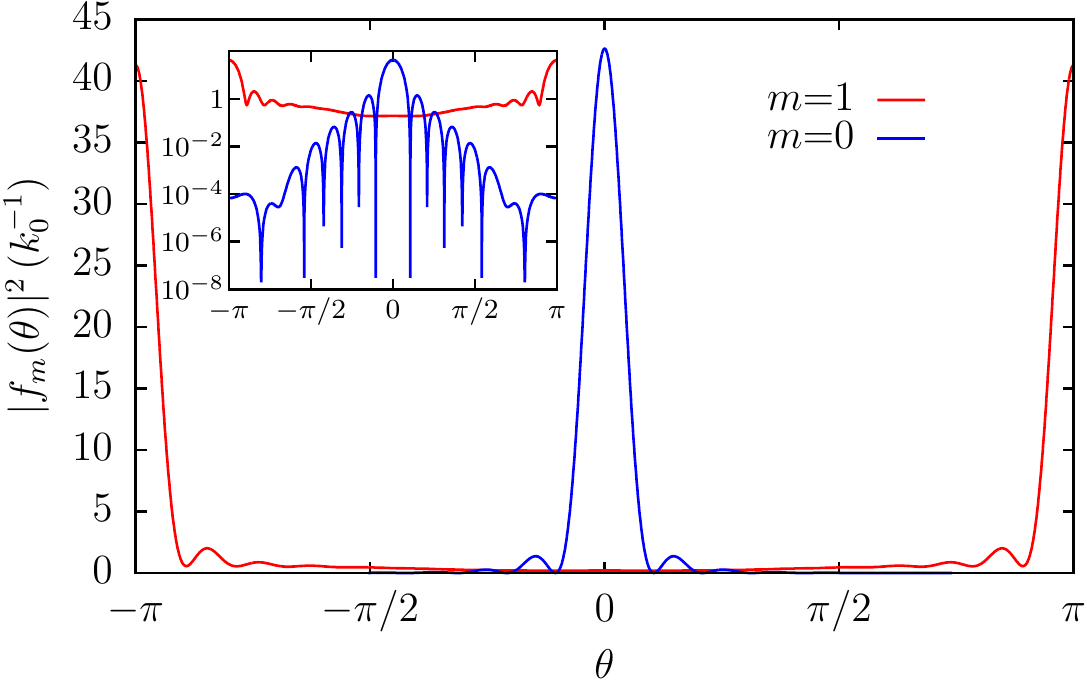}
\caption{(Color online) (a) Differential cross section in every channel. Clearly, electrons scattered elastically ($m=0$ channel) move mainly forward ($\theta=0$), while those that are scattered inelastically ($m=1$ channel) are backscattered ($|\theta|=\pi$). The inset in logarithmic scale shows that there are directions where the elastically scattered wave is suppressed ($f_0(\theta)=0$), which might correspond to destructive interference due to internal reflections. Here we  used $\eta=0.23$, $R=10k_0^{-1}$ and $\mu=0$ \label{figure_bound_states}}
\end{center}
\end{figure}
%%%%%%%%%%%%%%%%%%%%%%%%%%%%%%%%%%%%%%%%%%%%%%%%%%%%%%%%%%

Figure \ref{figure_bound_states} shows the differential cross sections as a function of $\theta$ for $\mu=0$. These are peaked at $\theta=0$ ($\pi$) for $m=0$ ($1$) and symmetrical around $\theta=0$. In the $m=0$ channel electrons are mainly transmitted forward while the backscattering takes place (as before) in the $m=1$ channel. The inset shows the same data in logscale, in order to highlight the zeros in $|f_0(\theta)|^2$. These zeros   signal the destructive interference between different $r_0^{(l)}$ modes and the incident wave.  The differential cross section $|f_1(\theta)|^2$ does not exhibit any zero. 

%%%%%%%%%%%%%%%%%%%%%%%%%%%%%%%%%%%%%%%%%%%%%%%%%%%%%%%%%%
\subsubsection*{Near field scattering}
\begin{figure}[t]
\begin{center}
\includegraphics[width=0.95\columnwidth]{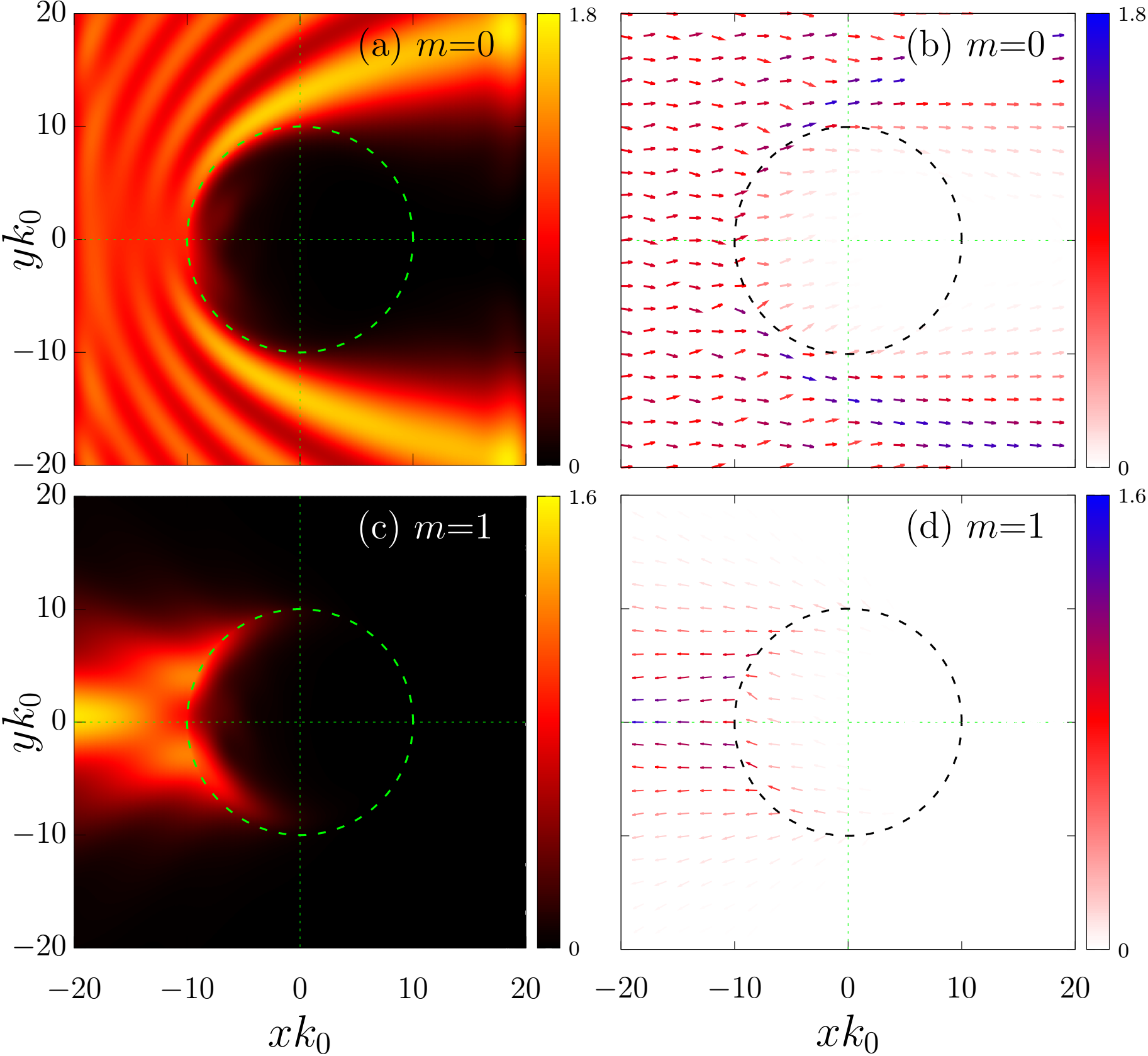}
\caption{(Color online) Density of probability [(a)] and current density [(b)] plots for the channel $m=0$ (including both the incident wave and the scattered one). In (a) we see a clear depletion (shadow) in the constant background of the incident plane wave, as expected by the interference of the incident and (forward) scattered waves. (c) and (d), the same in channel $m=1$.  In (b)  we see an accumulation of probability to the left the circular spot, making clear the presence of backscattering in this channel.  In both channels it is evident a lack of symmetry around $\theta=\pi$ due to the presence of a chiral current inside the irradiated region. The parameters used here are $k_0R=10$, $\mu=0$ and $\eta=0.2$ and the dashed circles indicate the border of the irradiated region \label{plane_wave_density_map}.}
\end{center}
\end{figure}

%%%%%%%%%%%%%%%%%%%%%%%%%%%%%%%%%%%%%%%%%%%%%%%%%%%%%%%%%%
To analyze the scattering near the irradiated spot we calculate all the coefficients that defines the Floquet wave function inside and outside the irradiated region and evaluate both the probability density and the probability current. We only discuss the case $\mu=0$ (center of the dynamical gap) for the sake of simplicity. As in the previous cases, we expect both magnitudes to be sensitivity to the helicity of the vector potential field. To this end we plot in Figs.~\ref{plane_wave_density_map}(a) and \ref{plane_wave_density_map}(b) a color map of the probability density $|\Phi_0(r,\theta)|^2$ and the vector field of the probability current, $\bm{J}_0=\Phi_0^\dagger\bm{\sigma}\Phi_0$, respectively, for the $m=0$ channel. The corresponding plots for the $m=1$ channels are presented in Figs.~\ref{plane_wave_density_map}(c) and \ref{plane_wave_density_map}(d). In the $m=0$ channel we plot the complete wave function (incident plus reflected) in order to expose the depletion (shadow) to the right of the spot  in the constant background given by the incident wave. In a clear contrast with the far field limit where the differential cross section is symmetric around $\theta=\pi$, here there is a clear asymmetry, which we interpret as a manifestation of the Goos-H\"anchen shift discussed previously---a quantitative discussion is given in the next section. Notice that inside the irradiated spot there is a clear chiral current in the (clockwise) $-\hat{\bm{\theta}}$ direction. The presence of a GH shift is possible here for a plane wave by virtue of the circular geometry of the interface.
%%%%%%%%%%%%%%%%%%%%%%%%%%%%%%%%%%%%%%%%%%%%%%%%%%%%%%%%%%
%%%%%%%%%%%%%%%%%%%%%%%%%%%%%%%%%%%%%%%%%%%%%%%%%%
%%%%%%%%%%%%%%%%%%%%%%%%%%%%%%%%%%%%%%%%%%%%%%%%%%%%%%%%%%
\begin{figure}[t]
\begin{center}
\includegraphics[width=0.95\columnwidth]{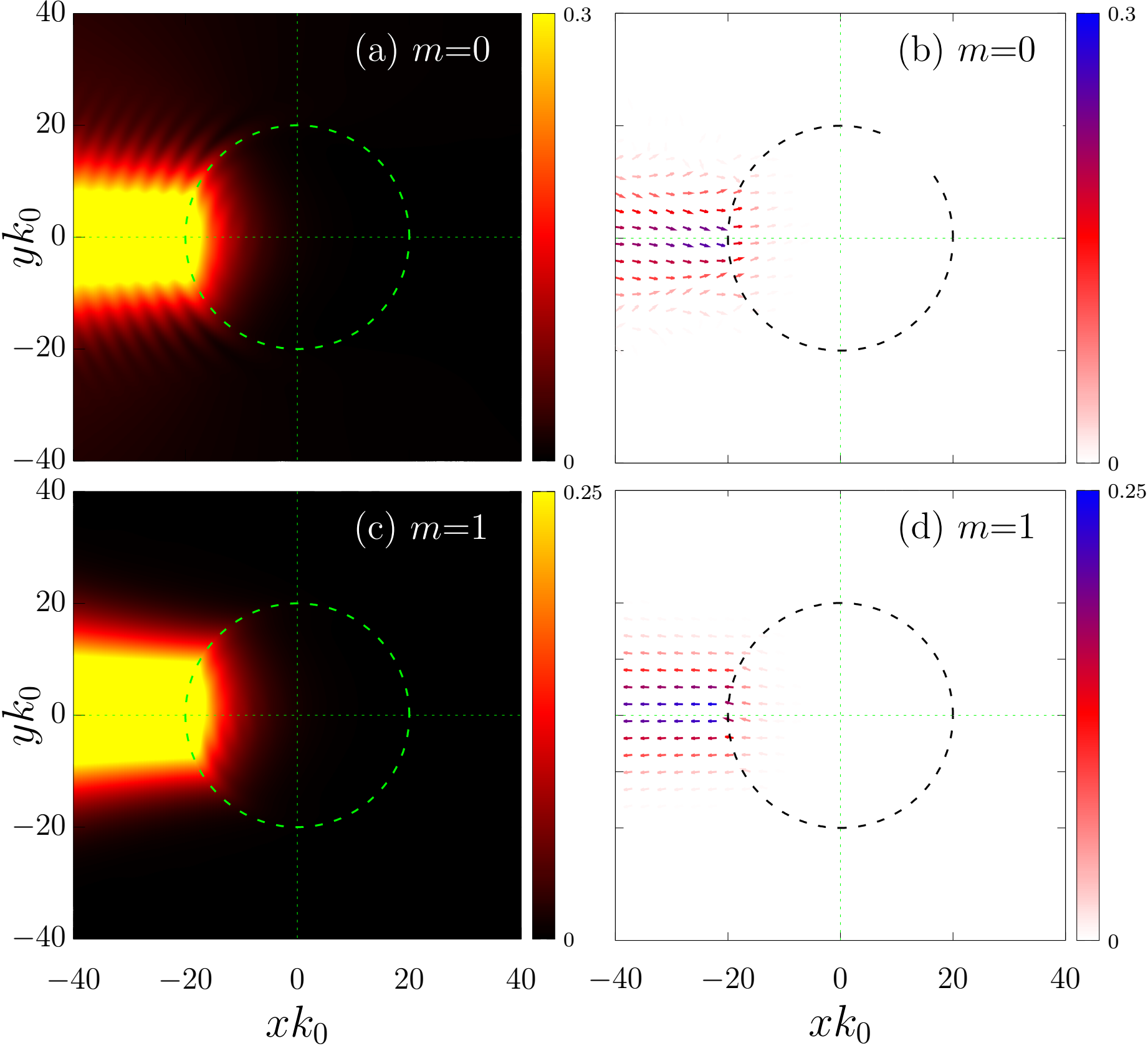}
\caption{(Color online) Same as the previous figure but for an incident beam (finite size cross section) and $k_0R=20$. The asymmetry around $\theta=\pi$ is apparent as well as the chiral character of the current inside the irradiated spot.  \label{density_beam}}
\end{center}
\end{figure}
%%%%%%%%%%%%%%%%%%%%%%%%%%%%%%%%%%%%%%%%%%%%%%%%%%
%%%%%%%%%%%%%%%%%%%%%%%%%%%%%%%%%%%%%%%%%%%%%%%%%%%%%%%%%%
\subsection{A finite size beam}
\begin{figure}[t]
\begin{center}
\includegraphics[width=0.85\columnwidth]{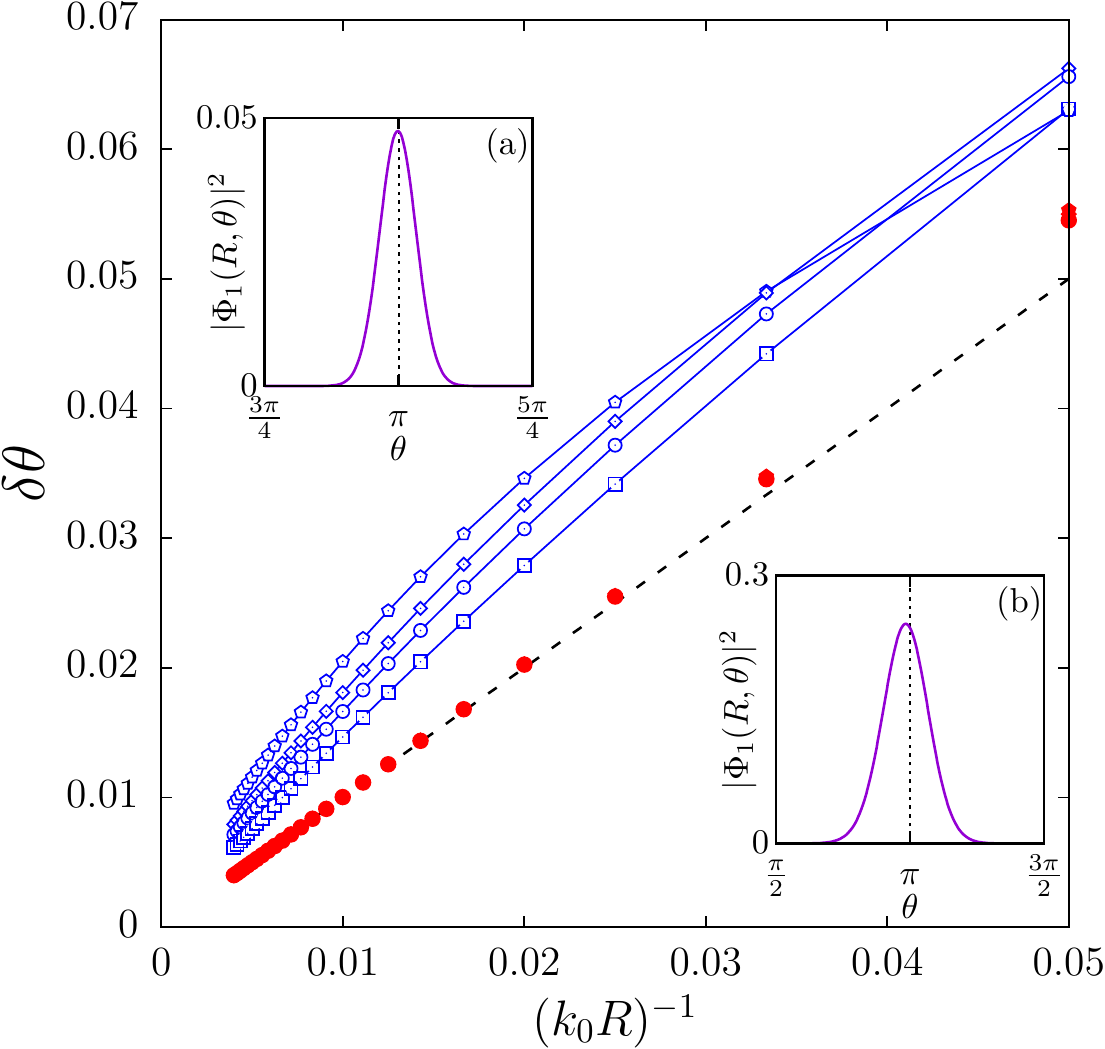}
\caption{(Color online) Angular Goos-H\"anchen shift $\delta\theta$ as a function of the inverse of $k_0R$ for the wave backscattered in channel $m=1$. Full (open) symbols correspond to an incident beam (plan wave) for $\eta=0.1$ (pentagon), $0.2$ (diamond), $0.3$ (circle), $0.6$ (square). The dashed line corresponds to $\delta\theta=1/k_0R$ (see text). 
The insets show $|\Phi_1(R,\theta)|^2$ for $k_0R=250$ [(a)] and $k_0R=20$ [(b)]. \label{goos2}.}
\end{center}
\end{figure}

%%%%%%%%%%%%%%%%%%%%%%%%%%%%%%%%%%%%%%%%%%%%%%%%%%%%%%%%%%
To better identify the GH shift, we now consider an incident beam, that is a wave front with a finite cross section. This is done in the same way as for the planar interface by constructing a superposition of plane waves (cf. Eq.~\eqref{phi0}). 
Figure \ref{density_beam} shows the results where the asymmetric distribution around $\theta=\pi$ is apparent. In order to quantify it we calculated the average angular shift at $r=R$. Namely,
%%%%%%%%%%%%%%%%%%
\begin{equation}
\delta\theta=\frac{\int_{-\pi}^\pi  (\pi-\theta) \,|\Phi_1(R,\theta)|^2\,d\theta}{\int_{-\pi}^\pi  |\Phi_1(R,\theta)|^2\,d\theta}\,.
\end{equation}
%%%%%%%%%%%%%%%%%%
Figure \ref{goos2} shows $\delta\theta$ as a function of the inverse of $k_0R$ for both the incident beam (solid red symbols) and the incident plane wave (open blue symbols). Two profiles of $|\Phi_1(R,\theta)|^2$ are also shown for two different values of $k_0R$. While the overall trend is similar, the shift for the two cases present different features. On the one hand, for the plane wave the shift is larger and sensitive to the value of $\eta$, being smaller the larger $\eta$ is and clearly different from the linear behavior we found in Sec.~\ref{goosH}. On the other hand,  the shift in the beam case presents a clear linear behavior for large $k_0R$ while it is rather insensitive to the value of $\eta$ (the different overlapping  types of solid symbols correspond to different values of $\eta$). This is consistent with an origin of the asymmetry in the anomalous GH shift found in previous sections:  
for normal incidence in a planar geometry we know that $\delta_{r1}=k_0^{-1}$, and since we can relate it with the angular shift (for small $\delta\theta$) as $\delta_{r1}{\approx}R\,\delta \theta$, we get
%%%%%%%%%%%%%%%%%%%%%%%%%%%%%%%%%%%%%%%%%%%%%%%%%%%%%%%%%%
\begin{equation}
  \delta \theta \approx \frac{1}{k_0R}.
\end{equation}
%%%%%%%%%%%%%%%%%%%%%%%%%%%%%%%%%%%%%%%%%%%%%%%%%%%%%%%%%%
The exact linear dependence with the inverse of $k_0R$ is shown in Fig.~\ref{goos2} with a dashed black line. It must be stressed that this behavior is expected when the width of the beam is small in comparison with the diameter of the irradiated spot, roughly $\sigma^{-1}\gg R$, where $\sigma$ is the width of the Gaussian beam in momentum coordinates [cf. Eq.~\eqref{phi0} and the definition of $f(k_y-\bar{k}_y)$]. This explains the deviations from this trend for large values of $(k_0R)^{-1}$. 
%Moreover, this distinction cannot be made for the plane wave since this can be understood as an {\it infinitely broad} beam, whose width is always much larger that the size of the spot. This explains why the parallel with the straight interface fails here and we cannot approach the linear dependency $\delta\theta=1/(k_0R)$.
%%%%%%%%%%%%%%%%%%%%%%%%%%%%%%%%%%%%%%%%%%%%%%%%%%%%%%%%%%
%%%%%%%%%%%%%%%%%%%%%%%%%%%%%%%%

%%%%%%%%%%%%%%%%%%%%%%%%%%%%%%%%
\section{Final remarks}
%%%%%%%%%%%%%%%%%%%%%%%%%%%%%%%%
We have presented a throughout discussion of the inelastic scattering  of Dirac fermions induced by the presence of a circularly polarized electromagnetic field in a given region. The analyses was carried out using the Floquet formalism that allows to treat the problem as a multichannel  scattering. 
As we only consider the case where the irradiated  region contains evanescent modes (incident energies inside the dynamical gap, $\varepsilon\sim\hbar\Omega/2$)  the incident wave must be fully reflected. Own to the fact that the perturbation is circularly polarized, the reflected wave must have its pseudospin flipped, so that the reflection occurs mainly on the $m=1$ channel (inelastic scattering).  Furthermore, we retained only two Floquet replicas ($m=0$ and $m=1$ channels) as the time dependent field was assumed to be small ($\eta\ll1$)---while this is a significant simplification for the analytical treatment, it still allows to obtain some general results valid even when including other replicas, provided $\eta$ remains small. 

We found that this scattering at a region with broken time reversal symmetry leads to the appearance of an anomalous GH shift both for the planar and the circular geometry. The GH shift in the inelastic channel ($m=1$) turns out to be anomalous, in the sense that its sign does not depend on the incident angle of the beam but on the helicity of the circularly polarized field in the irradiated region. In addition, its value is universal (for $\varepsilon=\hbar\Omega/2$) as it does not depend on the intensity $\eta$ of the field.

Quite notably, the presence of such a shift can be related to existence of topological edge states at the interface between two irradiated regions with opposite polarizations. 
From this perspective, if we consider a finite narrow channel between the two irradiated regions, as shown in Fig.~\ref{ES}, the GH shift could be related to transport properties along the channel. This remains an interesting prospect for future investigations.

%%%%%%%%%%%%%%%%%%%%%%%%%%%%%%%%%%% 
\begin{acknowledgments}
We acknowledge financial support from ANPCyT (grants PICTs 2013-1045 and 2016-0791), from CONICET (grant PIP 11220150100506) and from SeCyT-UNCuyo (grant 06/C526).
\end{acknowledgments}
%%%%%%%%%%%%%%%%%%%%%%%%%%%%%%%%%%%


%\bibliographystyle{apsrev4-1_title}
%\bibliography{GH}
%\input{Goos-Hanchen_v8.bbl}

\end{document}